\documentclass[conference]{IEEEtran}
\IEEEoverridecommandlockouts
\usepackage{cite}
\usepackage{amsmath,amssymb,amsfonts}
\usepackage{algorithmic}
\usepackage{graphicx}
\usepackage{textcomp}
\usepackage{xcolor}
\def\BibTeX{{\rm B\kern-.05em{\sc i\kern-.025em b}\kern-.08em
    T\kern-.1667em\lower.7ex\hbox{E}\kern-.125emX}}

\usepackage{xspace}
\usepackage{pifont}
\usepackage{algorithm}
\usepackage{subcaption}

\usepackage[bookmarks=true,breaklinks=true,letterpaper=true,colorlinks,linkcolor=blue,citecolor=blue,urlcolor=black]{hyperref}

\newcommand{\old}[1]{}
\newcommand{\fig}[1]{Figure~\ref{#1}}
\newcommand{\sect}[1]{Section~\ref{#1}}
\newcommand{\tab}[1]{Table~\ref{#1}}
\newcommand{\algo}[1]{Algorithm~\ref{#1}}

\newcommand{\proposedname}[0]{PIM-MMU\xspace}
\newcommand{\pluseq}{\mathrel{+}=}

\newcommand{\INDSTATE}[1][1]{\STATE\hspace{#1\algorithmicindent}}

\begin{document}

\title{\proposedname: A Memory Management Unit for Accelerating 
	Data Transfers in \\Commercial PIM Systems
 \thanks{
\noindent\rule{4cm}{0.4pt}\newline
This is the author preprint version of the work. The authoritative version will appear in the Proceedings of the 57th IEEE/ACM International Symposium on Microarchitecture (MICRO-57), 2024.
}
 }

\author{
\IEEEauthorblockN{Dongjae Lee \hspace{1em} 
                  Bongjoon Hyun \hspace{1em} 
                  Taehun Kim \hspace{1em} 
                  Minsoo Rhu}
\vspace{1em}
\IEEEauthorblockA{School of Electrical Engineering\\KAIST
\\\texttt{\{dongjae.lee, bongjoon.hyun, taehun.kim, mrhu\}@kaist.ac.kr}}
\\
}
\linespread{0.985}

\maketitle
\begin{abstract}

Processing-in-memory (PIM) has emerged as a promising solution for accelerating
memory-intensive workloads as they provide high memory bandwidth to the
processing units. This approach has drawn attention not only from the academic
community but also from the industry, leading to the development of real-world
commercial PIM devices. In this work, we first conduct an in-depth
characterization on UPMEM's general-purpose PIM system and analyze
the bottlenecks caused by the data transfers across the DRAM and PIM
address space.  Our characterization study
reveals several critical challenges associated with DRAM$\leftrightarrow$PIM
data transfers in memory bus integrated PIM systems, for instance, its high CPU core
utilization, high power consumption, and low read/write throughput 
for both DRAM and PIM. Driven by our key findings, we introduce
the \proposedname architecture which is a hardware/software co-design that enables
energy-efficient DRAM$\leftrightarrow$PIM transfers for PIM systems.  \proposedname
synergistically combines a hardware-based data copy engine, a
PIM-optimized memory scheduler, and a heterogeneity-aware memory mapping
function, the utilization of which is supported by our \proposedname software
stack, significantly improving the efficiency of DRAM$\leftrightarrow$PIM data
transfers.  Experimental results show that \proposedname improves the
DRAM$\leftrightarrow$PIM data transfer throughput by an average
$4.1\times$ and enhances its energy-efficiency by $4.1\times$,
	leading to a $2.2\times$ end-to-end speedup for real-world PIM
	workloads.

\end{abstract}

\begin{IEEEkeywords}
Processing-in-memory; near-memory processing; parallel architecture
\end{IEEEkeywords}

\section{Introduction}
\label{sect:introduction}

Modern data-intensive workloads (e.g., AI inference tasks for large language
		models~\cite{gpt2, gpt3, gpt4, opt, olive, attacc_cal, dfx, neupim}, recommendation
		systems~\cite{tensordimm, recsys, recnmp, tensorcast}, and graph
		processing~\cite{tesseract, graph_p, graph_h, graph_q, accelerating_gnn_on_pim}) are memory-bound as they
pose unprecedented demand for large data.  Despite the increasing demand for
high memory bandwidth, integrating a larger number of DRAM I/O pins at the
processor die is challenging because of form factor constraints and issues
related to signal integrity~\cite{aquabolt_xl_hbm2_pim_lpddr5_pim_with_in_memory_processing_and_axdimm_with_acceleration_buffer}. 

	To address such limitation, processing-in-memory (PIM) architectures gained interest by integrating compute logic close
	to DRAM. Because of its potential to alleviate the memory bandwidth
	bottleneck modern processors face, PIM has been extensively explored in both
	academia~\cite{a_logic_in_memory_computer, active_pages , chopim, computational_ram, gearbox, neurocube, pim_enabled_instructions, lazypim, to_pim_or_not, tesseract, ambit, spacea, simdram, conda, syncron, matsa, tom, damov,impica,accelerating_neural_network_inference_with_processing_in_dram, sequence_alignment_framework_using_pim} 
 and industry with several real-world PIM integrated systems introduced to
	the market~\cite{cxl_pnm,hc_aim_new,hbm_pim_isca,hc_upmem}.
	These PIM designs can be classified into two categories: (1) PIM
	integrated at the I/O bus (e.g., Samsung CXL-PNM~\cite{cxl_pnm} and SK Hynix
			AiMX~\cite{hc_aim_new}), and (2) PIM integrated at the host processor's
	memory bus (e.g.  UPMEM-PIM integrated with the CPU~\cite{hc_upmem}).
   In this work, we focus on the system-level
	challenges associated with memory bus integrated PIM architectures employing
	a \emph{bank-level} PIM core design (i.e., each memory bank contains a single PIM core)
	as they represent a state-of-the-art,
	commercially available PIM architecture, i.e., UPMEM-PIM~\cite{hc_upmem}.

	An important property memory bus integrated PIM commonly exhibits is
	its need to separate the address space of 
	DRAM vs. PIM (\sect{sect:background_pim_addr}).  Without such clear separation, 
	the host processor (whether it be CPU~\cite{hc_upmem} or
			GPU~\cite{hbm_pim_isca}) and PIM cores can simultaneously access the same
	memory bank within the PIM device,
	 leading to a structural hazard at shared
	resources (e.g., I/O bus within the PIM device). 
	Properly arbitrating host and PIM core's
	simultaneous memory accesses requires the host
	processor's memory controller to be heavily modified, making it 
	challenging to support such feature while still abiding by the strict latency constraints
	defined within DRAM protocols (e.g., DDR4).  Consequently,
	commercial PIM systems integrated at the memory bus circumvent this challenge by employing
	\emph{separate} physical address spaces for DRAM and PIM, allowing only a single
	entity (either the host processor or the PIM core) to access the PIM address
	space at any given time. As such, current PIM programming model requires programmers 
	to first allocate input
	data in the DRAM address space and then ``explicitly'' copy
	that data to the PIM address space when the PIM core is idle. 

In this work, we first uncover fundamental challenges associated with data
transfers across DRAM and PIM by characterizing UPMEM's commercial PIM
integrated system.  The data transfer
operations in UPMEM-PIM employ several optimization strategies including (1)
	the usage of AVX-512 vector load/store instructions~\cite{intel_avx} to
	migrate data across DRAM$\leftrightarrow$PIM, (2) which are heavily
	multi-threaded to maximize data transfer throughput over the off-chip memory
	channels.  Unfortunately, despite the heavy use of CPU cores and high 
	power consumption to explicitly
	orchestrate such data movements, we observe that the achieved data transfer
	throughput  is far from optimal e.g., only $11.6\%$ memory bandwidth
	utilization for DRAM reads and $15.5\%$ for PIM writes during
	DRAM$\rightarrow$PIM data copy operations, causing non-negligible performance
	overhead to the end-to-end program execution time. Through our detailed characterization and
	analysis, we identify the following two factors as the primary causes of suboptimal read
	(write) throughput from (to) \emph{both} PIM and DRAM:

\begin{itemize}

\item {\bf Low PIM read/write throughput due to software-based, coarse-grained
	memory scheduling.} In conventional memory systems, the memory mapping
	function that translates a physical address to a DRAM address partitions the
	data and distributes them across the DRAM subsystem in fine granularity to
	maximize memory-level parallelism (MLP) using channel/bank-group/bank-level
	parallelism.  Such fine-grained ``hardware-based'' memory mapping
	architecture (and the associated memory scheduling algorithm) is completely
	transparent to the software layer and helps improve memory bandwidth
	utilization. However, data transfers targeting the PIM address space cannot
	fully harness MLP because the data being transferred in and out of the PIM address
	space must be \emph{localized} to a specific memory bank (due to the
			bank-level PIM architecture
			design~\cite{hc_aim_new,hbm_pim_isca,hc_upmem}), rather than fine-grained
	interleaving them across the DRAM subsystem for maximum MLP.  Although the
	PIM runtime library attempts to better utilize MLP by employing software
	multi-threading (e.g., each thread handles data transfers targeting different banks
			within a given memory channel), we observe that such software-based
	coarse-grained memory scheduling falls short compared to conventional memory
	system's hardware-based fine-grained memory scheduling, leaving
	significant performance left on the table.

\item {\bf Low DRAM read/write throughput due to locality-centric (and not MLP-centric) 
	memory mapping.}
As mentioned above, memory bus integrated PIM systems separate the physical
address space of DRAM and PIM to obviate the need to modify the host
processor's memory controller, granting only a single entity (either the host
processor or a PIM core) to access PIM memory. This separation of DRAM vs.
PIM address space is implemented with a system BIOS update which employs a
memory mapping function that logically divides up the overall
physical address space into two mutually exclusive regions, one for DRAM and
the other for PIM.  We observe that such modification in memory mapping
throttles the MLP that normal DRAM read/write operations can reap out because
it ``homogeneously'' enforces a single, \emph{locality-centric} memory mapping function to both
DRAM and PIM physical addresses. Such design nullifies all the sophisticated MLP-enhancing features of conventional
memory mapping functions (e.g., XOR hashing), leading to aggravated DRAM
read/write throughput.

\end{itemize}

To this end, we propose a {\bf M}emory {\bf M}anagement {\bf U}nit for {\bf
	PIM} (\textbf{\proposedname}) which is designed to fundamentally address the
	challenges associated with DRAM$\leftrightarrow$PIM data transfers in memory bus
	integrated PIM system,
	synergistically combining the following three key components:

\begin{itemize}

\item {\bf Data Copy Engine.} In our proposed system, PIM programmers are
provided with a software interface that completely \emph{offloads}
DRAM$\leftrightarrow$PIM data transfers to a {\bf D}ata {\bf C}opy {\bf E}ngine
(DCE).  DCE not only handles the data copy operations but also the data
preprocessing operations (e.g., data transpose), accelerating the end-to-end
DRAM$\leftrightarrow$PIM data transfers without CPU intervention.

\item {\bf PIM-aware Memory Scheduler.} To overcome the limitations of PIM's
software-based coarse-grained memory scheduling, we propose a hardware-based,
{\bf PIM}-aware {\bf M}emory {\bf S}cheduler (PIM-MS) which is integrated
inside our DCE.  PIM-MS enhances PIM read/write throughput by leveraging the
unique properties of DRAM$\leftrightarrow$PIM data transfers where
coarse-grained data copy operations targeting different PIM cores (designated
		by PIM programmers at the software level) can be reordered without
affecting program correctness.  PIM-MS utilizes such property to enable
fine-grained memory scheduling at the hardware level, drastically improving MLP
and thus the PIM read/write throughput.

\item {\bf Heterogeneous Memory Mapping Unit.} We also introduce a unique
hardware-based memory mapping architecture named {\bf Het}erogeneous Memory {\bf Map}ping Unit (HetMap) that enables \proposedname 
to separate the
	address space of DRAM and PIM 
while also enabling high DRAM read/write throughput.
	HetMap maintains a dual set of memory mapping
	functions: (1) an \emph{MLP-centric} mapping function for memory transactions
	targeting the normal DRAM address space, and (2) a
	\emph{locality-centric} mapping function that is designed to honor the
	per-bank PIM address spaces by localizing the mapped regions within each PIM
	core's memory bank.

\end{itemize}

Putting everything together, \proposedname improves the DRAM$\leftrightarrow$PIM data
transfer throughput by an average $4.1\times$ (max $6.9\times$) 
	and enhances its energy-efficiency by
$4.1\times$ (max $6.9\times$), resulting in a $2.2\times$ 
end-to-end speedup (max $4.0\times$) for real-world PIM workloads.
\section{Background}
\label{sect:background}

\subsection{Memory Mapping Architecture}
\label{sect:background_mapping}

The memory mapping function within the memory controller is designed to map the
physical address space to the DRAM subsystem (i.e., channels, ranks, banks, and
		rows) while maximizing MLP. To enhance channel-level parallelism, for
instance, the memory mapping function maps memory channels using bits within
the physical address that change frequently. In general, bits closer to the
least significant bit (LSB) are likely to change more often, so the memory
mapping function tries to utilize these bits to distribute memory requests
evenly across the memory channels~\cite{gpu_mapping}. However, the frequency in
which bits within the address closer to the LSB change can vary depending
on the program's memory access pattern. To better accommodate such variability,
	 advanced memory mapping functions utilize XOR hashing~\cite{xor_mapping}.
	 Specifically, XOR hashing takes multiple physical address bits, from the LSB
	 to the most significant bit (MSB), to compute the channel address which
	 enables the DRAM subsystem to better adapt to diverse memory access
	 patterns.

\begin{figure}[t] \centering
  \includegraphics[width=0.49\textwidth]{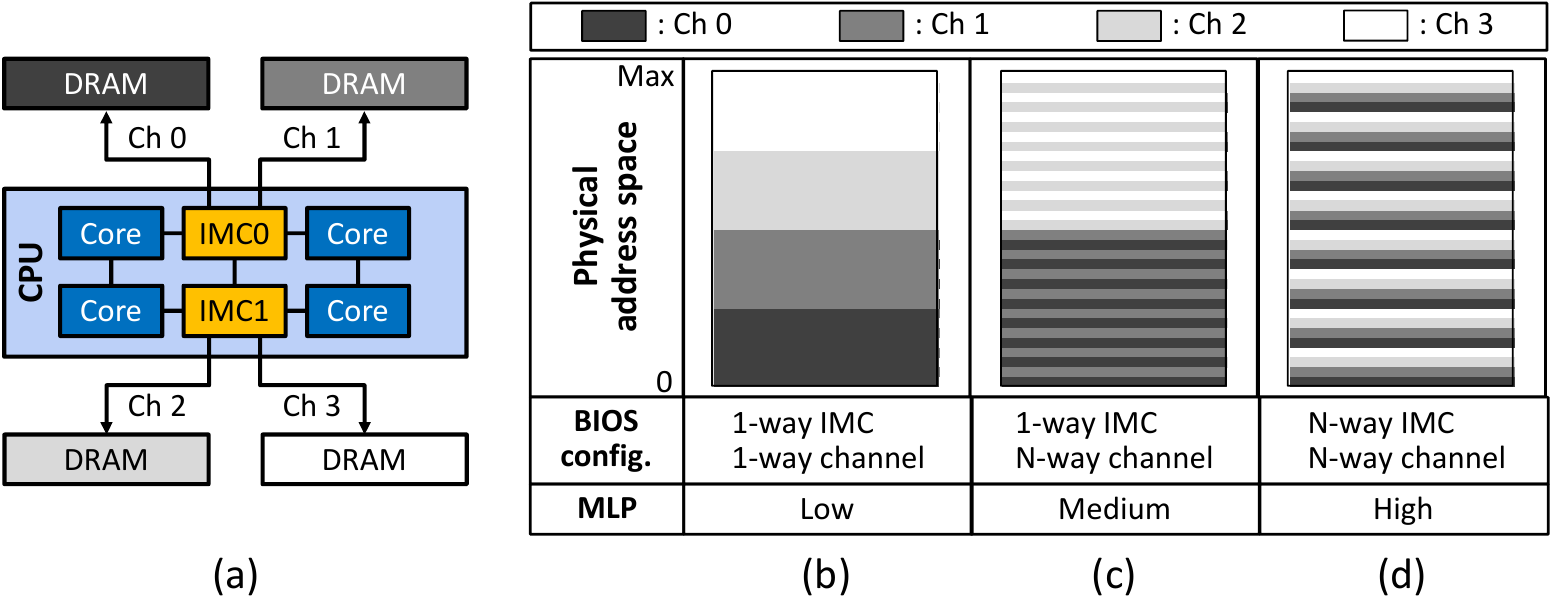} 
  \caption{(a) Intel Xeon CPU server's memory system topology. (b)-(d) show the BIOS configuration related to different memory mapping functions and how they translate into exploiting MLP. While the BIOS configuration also supports N-way NUMA/rank/bank-level interleaving, we omit discussing them for brevity.}
  \label{fig:intel_mapping}
	\vspace{-1.1em}
\end{figure}

In modern high-end server class x86 CPUs (or GPUs), the memory mapping function
can be customized by adjusting the BIOS configuration (or vBIOS in the context
		of GPUs~\cite{sait_pimlibrary}). In Intel Xeon CPUs, for instance, the BIOS
configuration provides a knob to enable (aka N-way) or disable (aka 1-way)
	address interleaving across different hierarchical levels of the DRAM
	subsystem (\fig{fig:intel_mapping}). Turning on N-way interleaving that targets a
	specific DRAM subsystem (e.g., Integrated Memory Controller (IMC) level
			interleaving, channel level interleaving, $\ldots$) thus 
	enables the memory mapping function to exploit MLP at that particular DRAM
	subsystem.  As shown in \fig{fig:intel_mapping}(b), configuring the memory
	mapping function as 1-way interleaving for \emph{both} IMC level and channel
	level positions their corresponding address bits
	closer to the MSB, making it challenging to fully exploit MLP.
	Conversely, in the configuration shown in \fig{fig:intel_mapping}(c),
	adjusting channel level interleaving to N-way moves the channel bits closer
	to the LSB, which helps better exploit MLP. However, maintaining the IMC
	level interleaving at 1-way results in the IMC bits to be placed near the
	MSB, rendering the lower physical address space to be mapped only at channels 0 and 1, connected to IMC0 (and the higher physical address space to be mapped 
				at channel 2 and 3, connected to IMC1).  To fully exploit MLP, the BIOS
			must be configured as N-way interleaving across \emph{all} DRAM subsystems which
			positions both the IMC and channel bits closer to the LSB
			(\fig{fig:intel_mapping}(d)). 

\subsection{PIM Integrated System and Its Address Space Management}
\label{sect:background_pim_addr}

Real-world industrial PIM devices such as HBM-PIM~\cite{hbm_pim_isca} and
UPMEM-PIM~\cite{hc_upmem} employ a \emph{bank-level} PIM architecture (one or two memory banks contain a single PIM core) which are integrated at the host processor's
memory bus alongside conventional DRAM.  Below we discuss the unique address
space management employed in memory bus integrated PIM.

\begin{figure}[t] \centering
  \includegraphics[width=0.49\textwidth]{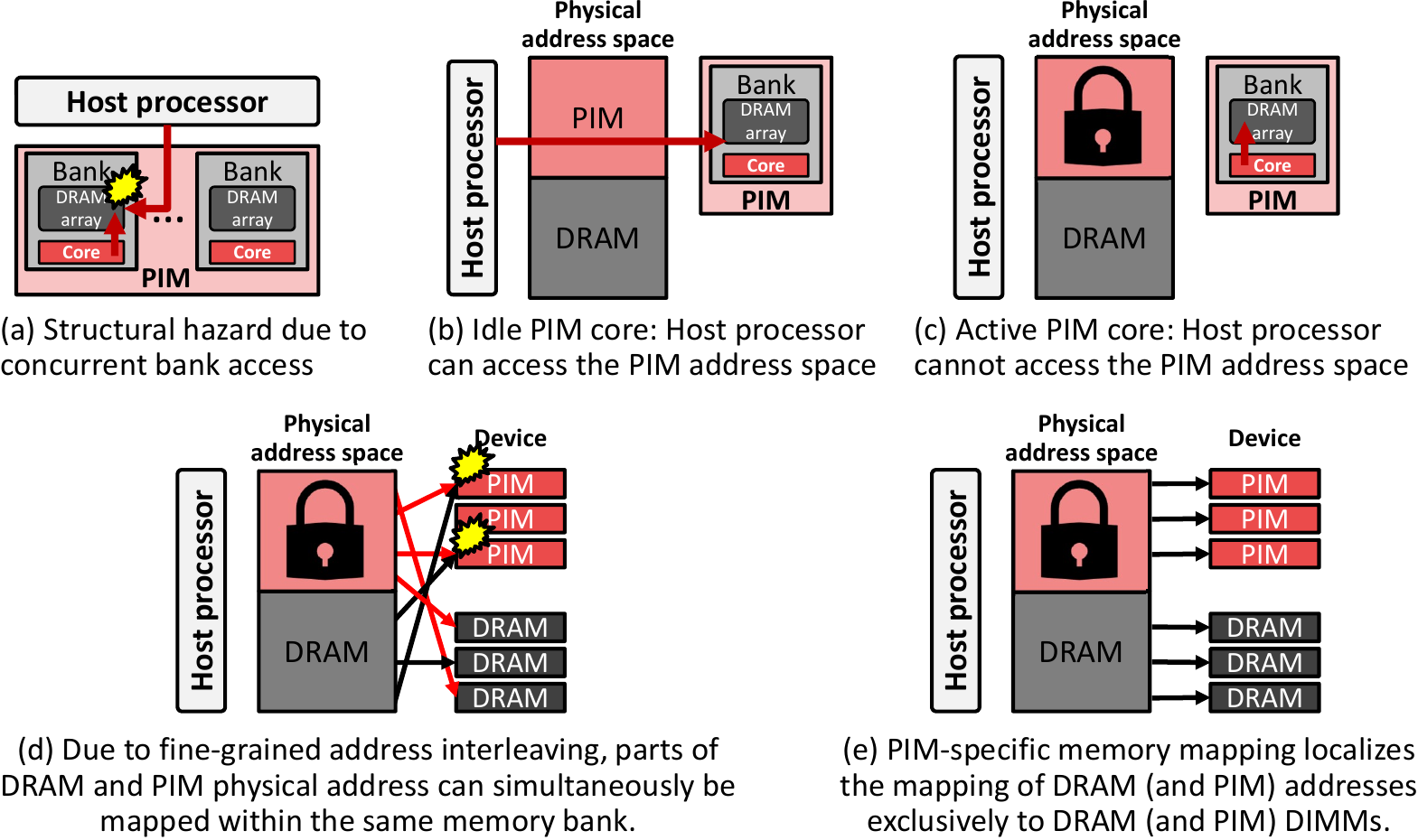} 
  \caption{Example showing how current PIM systems manage its DRAM and PIM physical address space.}
  \label{fig:pim_mapping_overview}
	\vspace{-1.3em}
\end{figure}

Current PIM systems employ a PIM-specific BIOS update to maintain
\emph{separate} physical address spaces for PIM and DRAM~\cite{upmem_sdk,hbm_pim_isca}.  This design decision
is due to the limitations coming from existing DRAM-specific protocols (e.g.,
		DDR4~\cite{ddr4_2400}), which dictate a deterministic latency behavior. To
better explain the intricacy of our problem in hand, consider the example in
\fig{fig:pim_mapping_overview}(a) where both the host processor and the PIM
core tries to access the same memory bank within the PIM device simultaneously.
Such situation complicates the task for the baseline host-side memory
controller in managing this structural hazard, unless it is heavily modified to properly
handle this conflicting scenarios. For instance, the memory controller would
have to arbitrate memory accesses between the host processor and PIM core,
		 which can lead to violation of the DRAM-specific protocols as their access
		 latency changes (i.e., arbitration cannot guarantee deterministic
				 latency).
Consequently, PIM manufacturers have designed their systems
		to prevent this situation from happening by employing the following two
		techniques. First, only a \emph{single} entity,
		either the host processor (\fig{fig:pim_mapping_overview}(b)) or the PIM core (\fig{fig:pim_mapping_overview}(c)), can access the PIM address space
		at any given time. This design decision obviates the need to modify the host processor's
		memory controller, easing PIM's integration in conventional systems. Second,
		 fine-grained address interleaving employed in conventional memory mapping (\fig{fig:intel_mapping}(d))
		is disabled so that it prevents segments of both the DRAM and PIM physical addresses from being  mapped
		to the same memory bank. As depicted in \fig{fig:pim_mapping_overview}(d), having parts of the DRAM and PIM physical address
		both be mapped at a common memory bank causes the resource conflicting scenario discussed in \fig{fig:pim_mapping_overview}(a).
		As such, the PIM-specific memory mapping in current PIM systems make sure that the physical DRAM (and PIM) addresses are
		mapped \emph{locally} within a DRAM (and PIM) DIMM (\fig{fig:pim_mapping_overview}(e)).
		A tradeoff
		made with this design is that the PIM programmer must \emph{explicitly} copy data
		across the DRAM address space and the PIM address space, whenever a data is
		offloaded from DRAM to PIM, and vice versa.

\subsection{UPMEM-PIM Hardware/Software Architecture}
\label{sect:background_hw_sw}

\textbf{Hardware architecture.} UPMEM-PIM is based on a DDR4-2400 DIMM form factor, equipped with eight UPMEM-PIM chips per rank.  Each UPMEM-PIM chip
contains eight PIM cores (called DPUs by UPMEM), one PIM core per each DRAM
bank.  A single host CPU can support up to 1,280 PIM cores and a
single PIM core is capable of achieving a peak memory bandwidth of 1 GB/s,
			 allowing the aggregate memory bandwidth to exceed 1 TB/sec.

\textbf{Programming model.} Similar to CUDA~\cite{cuda}, UPMEM-PIM adopts the
co-processor computing model, where the CPU offloads a memory-intensive task to
PIM. To offload a task to UPMEM-PIM, programmers are required to write two
distinct segments of code: the PIM-side code and the host-side code. In
the PIM-side code, the programmer describes the task to be offloaded to PIM
which follows the single-program multiple-data (SPMD) model, i.e., 
a single program gets executed by multiple PIM cores. Within the
	host-side code, the programmer designates the total number of PIM cores to
	utilize, which input data to transfer over to the PIM address space
	(DRAM$\rightarrow$PIM), and which output results derived by the PIM cores to
	transfer back into the DRAM address space (PIM$\rightarrow$DRAM), all of
	which is programmed using APIs provided in UPMEM-PIM's software
	stack.

\begin{figure}[t] \centering
  \includegraphics[width=0.49\textwidth]{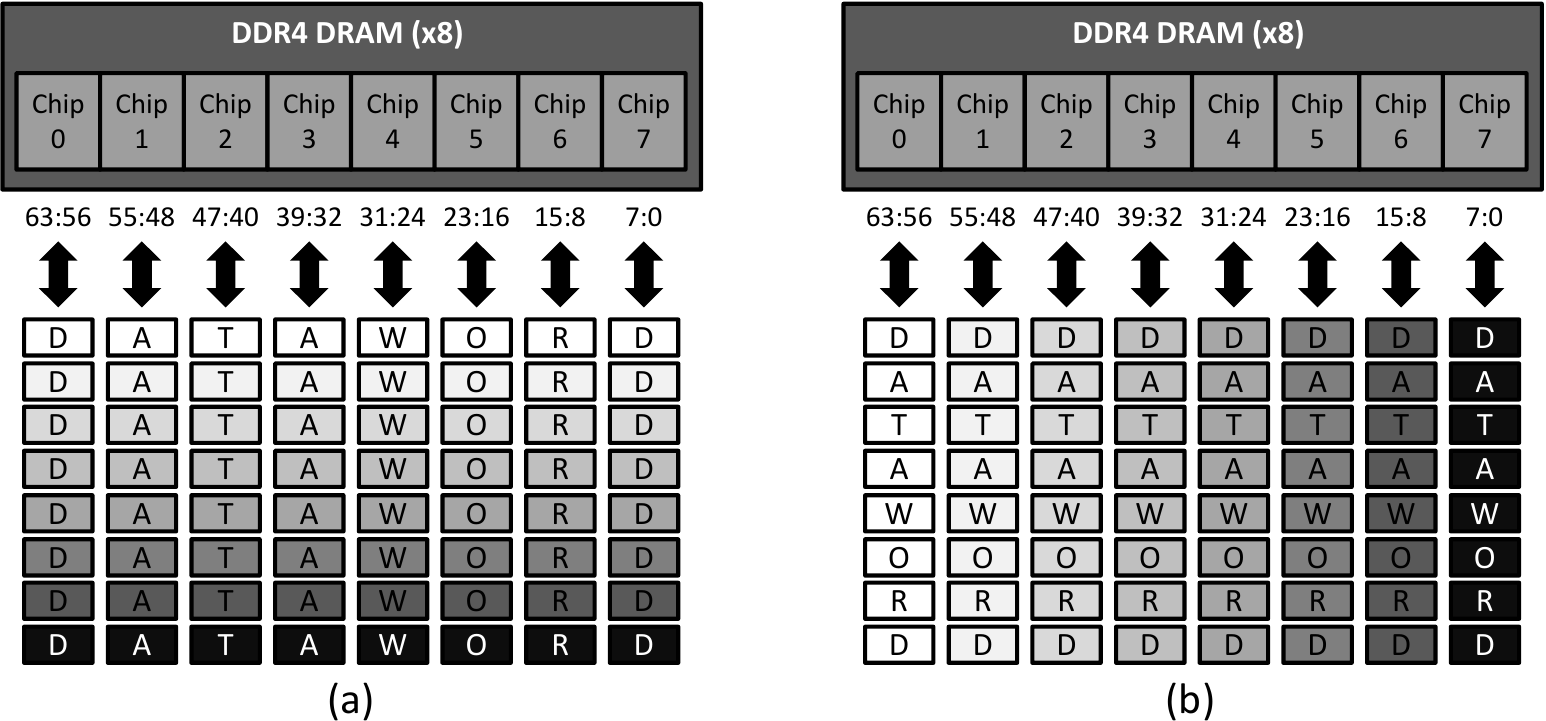} 
  \caption{Bytes that constitute a given data word (`D',`A',`T',`A',`W',`O',`R',`D') is colored identically. (a) Chip interleaving in a conventional DIMM-based memory system and (b) why UPMEM-PIM requires a transpose operation to be applied to the copied data beforehand to localize them within a single chip. }
  \label{fig:transpose_illustration}
	\vspace{-1em}
\end{figure}

\textbf{Runtime library for data transfers.} The UPMEM-PIM runtime library
offers a layer of abstraction that hides low-level details of the PIM hardware
from the programmer, one notable example being the need for
preprocessing data before they are transferred over to PIM. The need for data
preprocessing arises due to the way chip interleaving is employed within the
DIMM. Each data word ($8$ bytes) is partitioned in a 1-byte granularity
and distributed across multiple UPMEM-PIM chips ($8$ UPMEM-PIM chips in a $\times 8$
		configuration), an example we illustrate in
\fig{fig:transpose_illustration}(a). Such data interleaving across UPMEM-PIM chips
presents a significant
challenge for PIM computation because each (bank-level) PIM core only receives a fraction of a data word.
To address this issue, the UPMEM-PIM runtime library transposes the data into an (8$\times$8)
byte matrix and copies the transposed matrix across the 8 UPMEM-PIM chips, 
	allowing each PIM core to receive the full 8-byte data word within its own
	memory bank (\fig{fig:transpose_illustration}(b)). 

	When it comes to the actual DRAM$\leftrightarrow$PIM data transfer implementation (using \texttt{dpu\_push\_xfer}~\cite{upmem_sdk}),
	the runtime
	library employs several software optimizations to enhance
	the DRAM$\leftrightarrow$PIM data transfer throughput. These include 
	(1) the usage of AVX-512 \emph{vector} load/store instructions to transfer data in large
	chunks, and (2) using \emph{multi-threaded} implementations to initiate a large
	number of parallel data transfers concurrently as means to maximize data transfer throughput.
	Unfortunately, despite these efforts to
	optimize performance, our characterization reveals that the observed performance
	is far from ideal, which we root-cause in \sect{sect:motivation}.

\section{Motivation and System Characterization}
\label{sect:characterization}

\subsection{Motivation}
\label{sect:motivation}

This paper explores the system-level challenges associated with memory bus
integrated PIM systems employing a \emph{bank-level} PIM architecture as they represent a state-of-the-art, commercially available PIM system, i.e.,
	UPMEM-PIM~\cite{hc_upmem}. In particular, we
	focus on UPMEM's general purpose PIM system due to their immediate market
	availability but more importantly their open-source software
	ecosystem driven by both industry~\cite{upmem, upmem_sdk}
	and academia~\cite{pimulator,prim,prim_2, prim_3, upmem_sigmod, a_case_study_of_processing_in_memory_in_off_the_shelf_systems, trans_pim_lib, trans_pim_lib_arxiv, simplepim, 
 upmem_ispass_ml, upmem_isvlsi_ml, upmem_arxiv_ml, sparsep, sparsep_arxiv}. 

This section conducts a characterization on
	UPMEM-PIM to root-cause the underlying challenges of its
	DRAM$\leftrightarrow$PIM data transfers (\sect{sect:methodology} details our evaluation methodology). 
Memory bus integrated PIM systems
employ separate physical addresses for DRAM and PIM, necessitating explicit
DRAM$\leftrightarrow$PIM data transfers (\sect{sect:background_pim_addr}).
As we quantify in \sect{sect:evaluation}, the latency to transfer
data across these two regions incur significant performance overhead, accounting
to as much as $99.7\%$ (average $63.7\%$) of end-to-end execution time of our evaluated PIM workloads~\cite{prim}.
 Such
			observation is inline with prior work~\cite{prim,
				prim_2, prim_3, prim_repo,pimulator},
underscoring the importance of optimizing this critical system-level bottleneck.
	Next we root-cause
	the reason behind the sub-optimal performance of DRAM$\leftrightarrow$PIM data transfers.

\subsection{System Characterization and Key Challenges}
\label{sect:motivation_char}

\begin{figure}[t] \centering
  \includegraphics[width=0.49\textwidth]{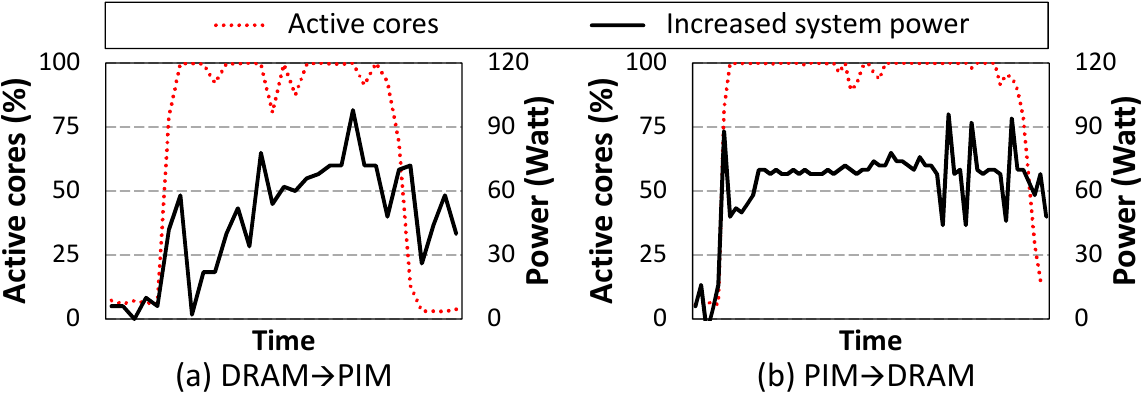} 
  \caption{The fraction of active CPU cores (left axis) and system power consumption (right axis) during 
		(a) DRAM$\rightarrow$PIM and (b) PIM$\rightarrow$DRAM data transfer. System power
			consumption	is measured using Intel's Performance Counter Monitor (PCM)~\cite{intel_pcm}.}
  \label{fig:pwr_consumption}
	\vspace{-1.3em}
\end{figure}

{\bf (Challenge \#1) High CPU core utilization and power consumption.} The
DRAM$\rightarrow$PIM data transfer involves three main stages: (1) reading from
DRAM, (2) preprocessing, (3) and writing to PIM, with the reverse sequence
applied for PIM$\rightarrow$DRAM transfers. Our first key observation is
that, because the CPU is in charge of orchestrating the entire process of data
transfers, it significantly taxes the host processor, leading to high power
overheads.  \fig{fig:pwr_consumption} illustrates the effect of
DRAM$\leftrightarrow$PIM data transfers on CPU core utilization and system
power consumption.  As discussed in \sect{sect:background_hw_sw}, the data
transfer operation in the UPMEM-PIM runtime library is implemented using
AVX-512 vector load/store instructions,	 which are known to be power
hungry~\cite{avx512_pwr1, avx512_pwr2}.  Consequently, data transfers across
DRAM vs. PIM addresses push CPU core utilization to near maximum levels and
reach close to $70$ Watts of system power consumption.

		Despite such high power overheads, DRAM$\leftrightarrow$PIM data transfers
		exhibit limited efficiency from a throughput perspective, significantly
		underutilizing available memory bandwidth.
 Below we
		root-cause the reason behind \emph{why} contemporary PIM system achieves such low data
		transfer throughput.

\begin{figure}[t] \centering
  \includegraphics[width=0.495\textwidth]{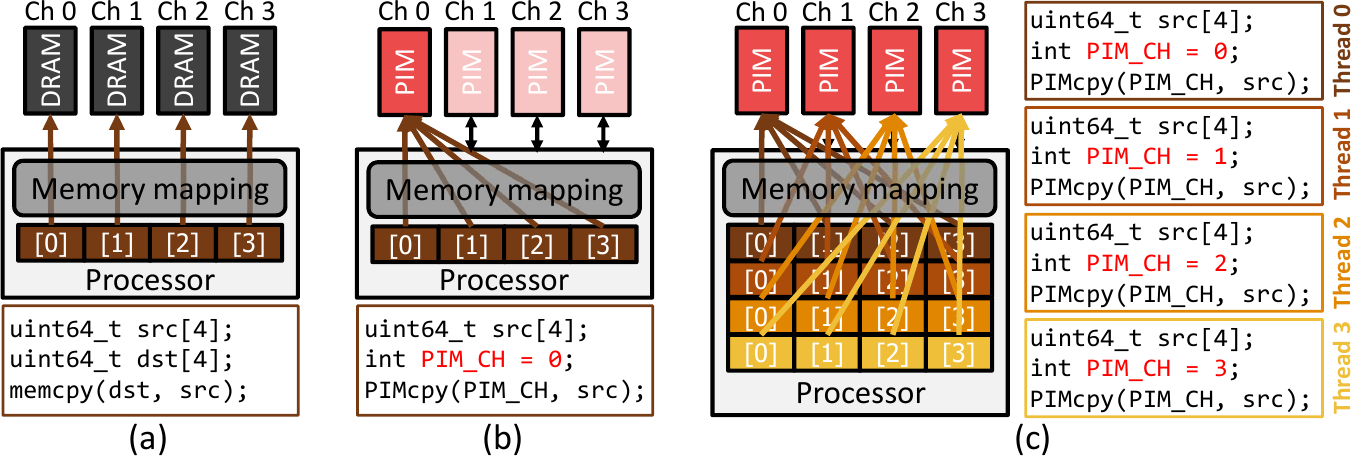} 
  \caption{(a) MLP-optimized, hardware-based data transfer that maximally utilizes
		memory bandwidth. (b) Software-based data transfer using a \emph{single} PIM
			thread which transfers data
			to its designated memory bank, leading to underutilization of memory bandwidth.
			Such limitation is better addressed in (c) which utilizes \emph{multiple} concurrent PIM 
					 threads that target different channels/banks for data transfers,
				 achieving higher memory throughput. In the pseudo-code in (b-c), the role of \texttt{PIMcpy}
		is conceptually identical to UPMEM-PIM's \texttt{dpu\_push\_xfer} or CUDA's \texttt{cudaMemcpy} APIs.
}
  \label{fig:dram_vs_pim}
	\vspace{-1.3em}
\end{figure}

{\bf (Challenge \#2) Sub-optimal PIM read/write throughput.} 
Conventional DRAM
systems are provisioned with MLP-enhancing microarchitectural support. 
For instance, the memory mapping function, which translates
physical addresses to DRAM addresses, partitions/distributes the data
across the DRAM subsystem in fine granularity to maximize MLP.
Such ``hardware-based'' memory mapping is entirely transparent 
to the software layer and helps evenly distribute the memory read/write traffic across the memory channels (\fig{fig:dram_vs_pim}(a)).

In contrast, UPMEM-PIM is unable to fully reap out the MLP enhancing
opportunities inherent within such hardware-based memory mapping architectures
due to its \emph{bank-level} PIM design. One key characteristic of bank-level
PIM systems is their need for input data to be made \emph{locally} available
within the target PIM core's memory bank (\fig{fig:transpose_illustration}(b)),
			 before the PIM kernel is executed.  Consequently, PIM programmers must
			 explicitly designate which input data should be transferred over to
			 which PIM core's memory bank using software APIs that enable
			 DRAM$\leftrightarrow$PIM data transfers (e.g.,
					 \texttt{dpu\_push\_xfer} in UPMEM-PIM).  Unfortunately, because the transferred data is
			 targeted for a specific memory bank, it becomes challenging to
			 fully leverage channel/bank-group/bank-level parallelism in transferring
			 such data, leading to underutilization of PIM read/write throughput
			 (\fig{fig:dram_vs_pim}(b)).

\begin{figure}[t] \centering
  \includegraphics[width=0.45\textwidth]{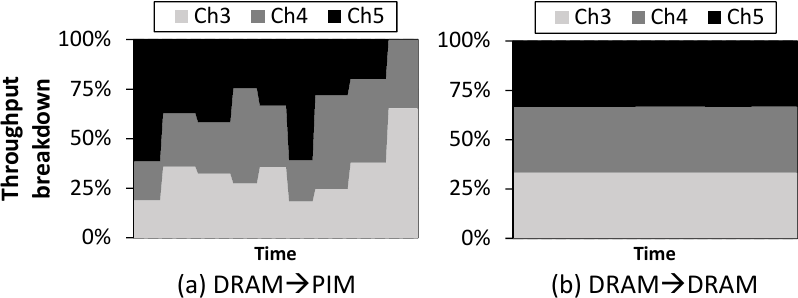} 
  \caption{Breakdown of data write throughput during (a) a software-based, coarse-grained DRAM$\rightarrow$PIM data transfer and (b) a hardware-based, fine-grained DRAM$\rightarrow$DRAM data transfer (i.e., \texttt{memcpy}). Examples assume that channels 0, 1, 2 are (read) source channels and channels 3, 4, 5 are (write) destination channels. We measure the per-channel write throughput using Intel VTune~\cite{intel_vtune} by executing microbenchmarks over real UPMEM-PIM systems. \sect{sect:methodology} details the microbenchmarks and our evaluated system configuration. Note that in (a) DRAM$\rightarrow$PIM data transfer, we manually assigned DRAM to channels 0-2 (read requests) and PIM to channels 3-5 (write requests) by adjusting the BIOS configuration. }
  \label{fig:bw_imbalance}
	\vspace{-1.3em}
\end{figure}

Given the limitation of bank-level PIM architectures in leveraging MLP, UPMEM-PIM
exploits \emph{thread-level parallelism} (TLP) to maximally utilize
available memory bandwidth.  As depicted in \fig{fig:dram_vs_pim}(c), the
UPMEM-PIM runtime library launches \emph{multiple} software threads for
DRAM$\leftrightarrow$PIM data transfers by having each thread to perform
read/write operations targeting different levels in the PIM hierarchy (e.g.,
		thread ID=0 targets PIM channel 0/bank 0 while thread ID=1 targets PIM
		channel 1/bank 0, $\ldots$), the goal of which is to maximize data
transfer throughput and MLP.  While utilizing TLP does help improve PIM
read/write throughput, our key observation is that such software-based
multi-threaded approach falls short in fully leveraging MLP for enhanced
performance. This is because the effectiveness of the multi-threaded PIM data
transfers is contingent upon how the Operating System (OS) schedules these
threads, which may not necessarily be aligned with the optimal distribution of
data accesses across the PIM hierarchy.  In general, the OS thread scheduling
policy prioritize fairness~\cite{linux_cfs} which may not necessarily lead to
threads being scheduled in a manner that balances its memory accesses across
the memory subsystem. This is natural as the OS is (and should be) unaware of
the existence of PIM.  For instance, if the OS thread scheduler prioritizes the
execution of three PIM threads transferring data in channel A and one
thread in channel B during a time interval of \texttt{T}, channel A will
experience a higher concentration of data accesses. To maintain system-wide
fairness, the scheduler may adjust the scheduling priority during the next time
interval of \texttt{T} to allow more threads to execute in channel B. While
such scheduling policy better guarantees fairness, it can lead to
inefficient utilization of memory bandwidth because certain memory channels are
preferentially accessed for longer periods of time, i.e., time interval
\texttt{T} is in the order of several $ms$. In
\fig{fig:bw_imbalance}(a), we illustrate the implication of such
coarse-grained, software-based DRAM$\rightarrow$PIM data transfer. As shown,
	such software-based approach is not able to fully utilize MLP due to traffic
	congestion happening at certain PIM channels, unlike the conventional
	hardware-based data transfers (\fig{fig:bw_imbalance}(b)), which evenly
	distributes data traffic across all memory channels, leading to higher
	memory bandwidth utilization. Overall, data transfers targeting PIM exhibits
	sub-optimal memory bandwidth utilization, only achieving around $15.5\%$ of
	its theoretical peak value, e.g., averaging at 8.9 GB/sec vs. the maximum
	value of 57.6 GB/sec during DRAM$\rightarrow$PIM data transfer.	

\begin{figure}[t] \centering
  \includegraphics[width=0.49\textwidth]{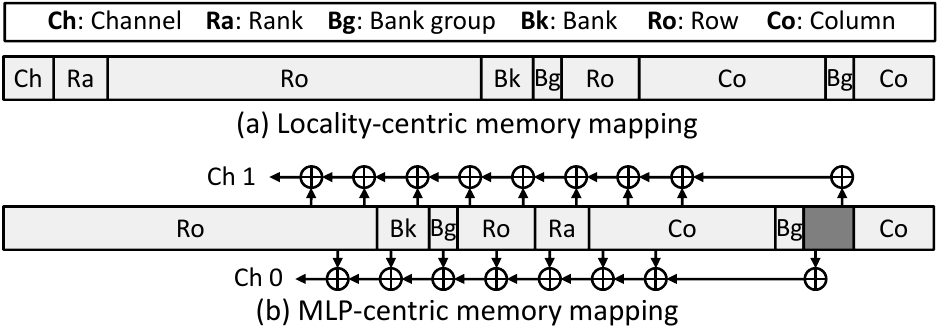} 
  \caption{(a) The locality-centric memory mapping employed in UPMEM-PIM  and (b) the MLP-centric memory mapping utilized in conventional systems without PIM, i.e., memory mapping (a) and (b) corresponds to those shown in \fig{fig:intel_mapping}(b) and \fig{fig:intel_mapping}(d), respectively. We referred to \cite{drama,intel_cascadelake_datasheet,edac} for the MLP-centric memory mapping of our baseline system without PIM.}
  \label{fig:pim_mapping_detail}
	%\vspace{-1.3em}
\end{figure}

\begin{figure}[t] \centering
  \includegraphics[width=0.485\textwidth]{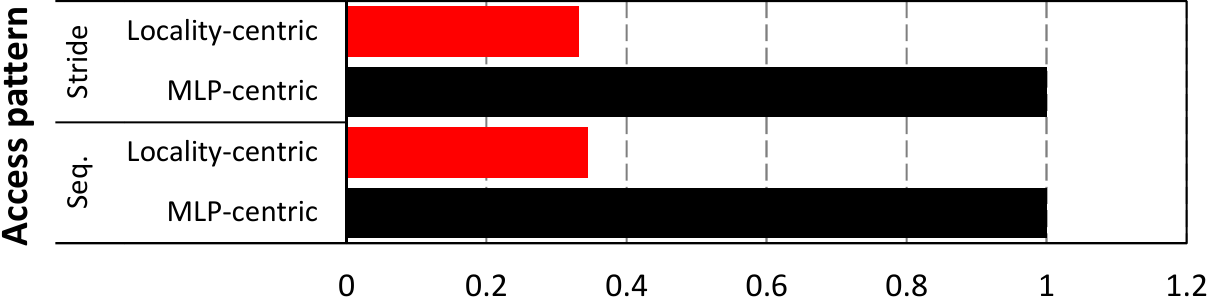} 
  \caption{Normalized DRAM bandwidth utilization (x-axis) with a locality-centric mapping (red) and MLP-centric memory mapping (black) over sequential and strided memory access patterns.}
  \label{fig:pim_mapping_bw}
	\vspace{-1.3em}
\end{figure}

{\bf (Challenge \#3) Sub-optimal DRAM read/write throughput.} 
\sect{sect:background_pim_addr} discussed the need for separating the
physical addresses of DRAM vs. PIM using alternative memory mapping functions
via BIOS updates (\fig{fig:pim_mapping_overview}(e)).  An unfortunate side-effect
of such memory mapping is that it introduces a decrease in DRAM read/write
throughput.  We observe that the adjustment in memory mapping used for PIM
integrated systems removes MLP-optimized XOR hashing
techniques~\cite{xor_mapping} employed in conventional memory mapping functions.
Specifically, the adjusted memory mapping function
(\fig{fig:pim_mapping_detail}(a)) places the channel bits closer to the MSB to
\emph{localize} the mapping of PIM physical addresses to PIM DIMMs (and DRAM physical
		addresses to DRAM DIMMs), unlike the standard practice of positioning them
closer to the LSB while also employing XOR hashing to enhance MLP
(\fig{fig:pim_mapping_detail}(b)). Because only a \emph{single} memory mapping
function can be employed ``homogeneously'' within the overall memory system, \emph{both} PIM and DRAM
	DIMMs integrated at the memory bus are enforced with such \emph{locality-centric}
	memory mapping (\fig{fig:pim_mapping_detail}(a)), limiting the level of
	parallelism normal DRAM physical addresses can reap out\footnote{ It is worth
		clarifying that enabling MLP-enhancing address interleaving knobs (i.e.,
				N-way interleaving, see \fig{fig:intel_mapping}) only within the DRAM
			physical address space while disabling them for the PIM physical address
			space is not supported under the current system BIOS configuration.  }.
			In \fig{fig:pim_mapping_bw}, we compare the read/write
			throughput targeting normal DRAM physical addresses using locality-centric 
			memory mapping (red) vs. MLP-centric memory mapping (black). As depicted,
			the DRAM read/write throughput under the locality-centric
			 mapping is only 30\% of what is achievable with
			conventional MLP-centric mapping, regardless of the memory access pattern.
			This substantial throughput difference highlights the inability of PIM integrated system in fully exploiting MLP.
\section{\proposedname Architecture}
\label{sect:proposed}

\subsection{High-Level Overview}
\label{sect:proposed_overview}

\begin{figure}[t!] \centering
%\vspace{-1.3em}
\includegraphics[width=0.495\textwidth]{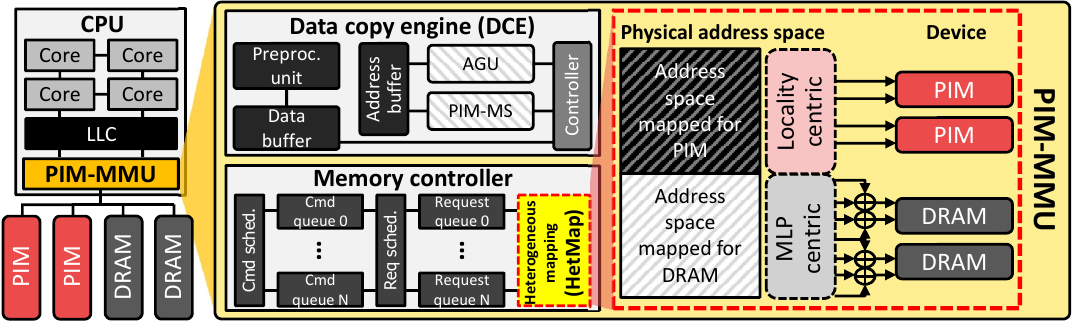} 
\caption{\proposedname architecture overview.}
	\vspace{-1em} 
	\label{fig:overview_proposed}
\end{figure}

Through our characterization in \sect{sect:characterization}, we root-caused the low
performance and high resource usage of DRAM$\leftrightarrow$PIM data transfers.
We propose a {\bf M}emory {\bf M}anagement {\bf U}nit for {\bf PIM}
(\proposedname), a hardware/software co-design that enables energy-efficient
DRAM$\leftrightarrow$PIM transfers for memory bus integrated PIM systems.
\fig{fig:overview_proposed} provides a high-level overview of \proposedname
which contains the following three key hardware components: (1) {\bf D}ata {\bf
	C}opy {\bf E}ngine (DCE), (2) {\bf PIM}-aware {\bf M}emory {\bf S}cheduler
	(PIM-MS), and (3) {\bf Het}erogeneous Memory {\bf Map}ping Unit (HetMap).
	User-level applications are able to utilize our \proposedname architecture
	via a dedicated software interface that completely \emph{offloads}
	DRAM$\leftrightarrow$PIM data transfers. In the remainder of this section,
	we first discuss \proposedname's  software
stack followed by a detailed description of
\proposedname's hardware architecture.

\subsection{Software Architecture}
\label{sect:proposed_sw}

\begin{figure}[t] \centering
  \includegraphics[width=0.45\textwidth]{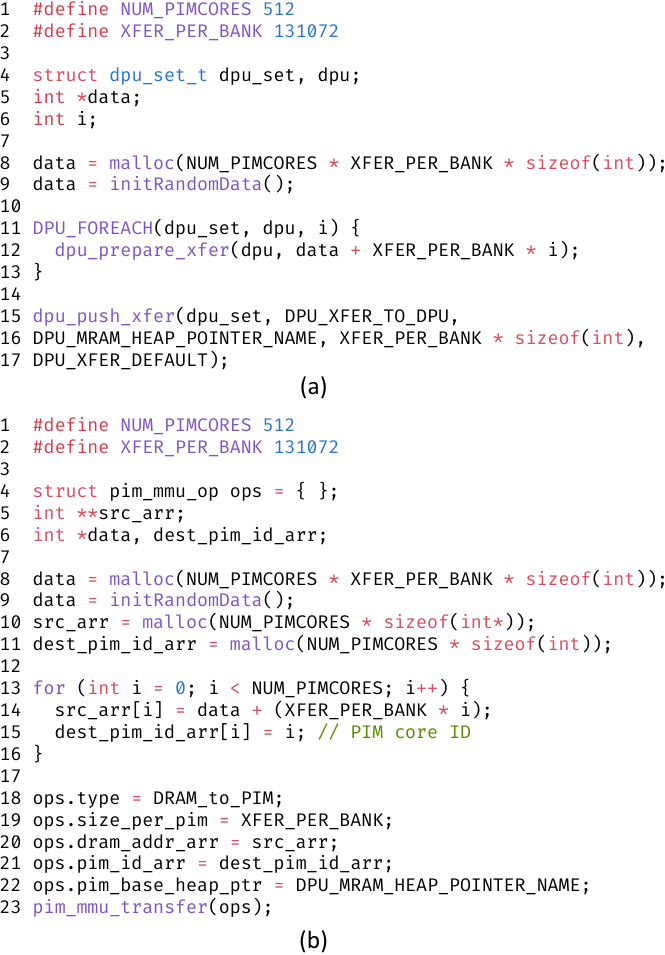} 
  \caption{Example pseudo-code showing how input data (128K elements for each PIM core) is
		transferred over to UPMEM-PIM using (a) conventional PIM programming APIs provided with UPMEM-PIM, and (b) 
			our proposed \proposedname specific APIs. It is worth pointing out that the PIM address (whether it be used
					as source or destination for data transfers) can be derived precisely using the PIM core ID (\texttt{dest\_pim\_id\_arr}) and
	the base heap pointer value (\texttt{DPU\_MRAM\_HEAP\_POINTER\_NAME})~\cite{upmem_sdk} (line $21$$-$$22$ in (b)).}
  \label{fig:code}
	\vspace{-1.3em}
\end{figure}

In our proposed system, PIM programmers are provided with the software
interface to leverage \proposedname for accelerating DRAM$\leftrightarrow$PIM
data transfers.  Specifically, our software stack contains the
following two components, the \proposedname runtime library and the
\proposedname device driver. We use the example in \fig{fig:code} that
conducts a DRAM$\rightarrow$PIM data transfer to
describe our software interface.  

{\bf User-level runtime library.}
The \proposedname runtime library offers a user-level API
(\texttt{pim\_mmu\_transfer}) that provides an abstraction to offload data
transfers to our hardware DCE.  This API utilizes a custom \texttt{struct} data
type (\texttt{pim\_mmu\_op}) as an input argument to acquire the necessary information
to offload DRAM$\leftrightarrow$PIM data transfers to the DCE (e.g.,
		data transfer direction (\texttt{DRAM\_TO\_PIM}), data transfer size per bank	(\texttt{XFER\_PER\_BANK}), and an array of
		pointers that designate where the source data (\texttt{src\_arr})
		as well as destination data (\texttt{dest\_pim\_id\_arr} and \texttt{DPU\_MRAM\_HEAP\_POINTER\_NAME}) are located (line $18$$-$$23$ in \fig{fig:code}(b)).
Unlike the baseline UPMEM-PIM implementation (\texttt{dpu\_push\_xfer}) where
\emph{multiple} threads orchestrate DRAM$\leftrightarrow$PIM data transfers
(line $11$$-$$15$ in \fig{fig:code}(a)), a call to \texttt{pim\_mmu\_transfer}
invokes a \emph{single} thread that offloads all the necessary information
required for DRAM$\leftrightarrow$PIM data transfers to the DCE.

{\bf Device driver.} The DCE is registered as an I/O device by mapping its corresponding Base Address Register (BAR) in the memory address space using MMIO (memory-mapped I/O). Existing on-chip components, such as the host processor's memory controller, already employ an MMIO-based approach for software-hardware communication\mbox{~\cite{edac}}, so our DCE can similarly be integrated into current software systems seamlessly. To support MMIO-based communication for the user-level \texttt{pim\_mmu\_transfer} API, the \mbox{\proposedname} device driver interacts with the  runtime library and manages MMIO access at the kernel-level. Specifically, when the \texttt{pim\_mmu\_transfer} API sends \texttt{pim\_mmu\_op} information to the \mbox{\proposedname} device driver, the driver writes this information to the MMIO region mapped to the DCE and finalizes the offloading of DRAM$\leftrightarrow$PIM data transfer, putting the requesting user process into sleep mode. Upon a successful data transfer completion, the \mbox{\proposedname} device driver receives an interrupt signal from the DCE, enabling the host processor to wake up and handle the interrupt appropriately.

\begin{figure}[t] \centering
  \includegraphics[width=0.485\textwidth]{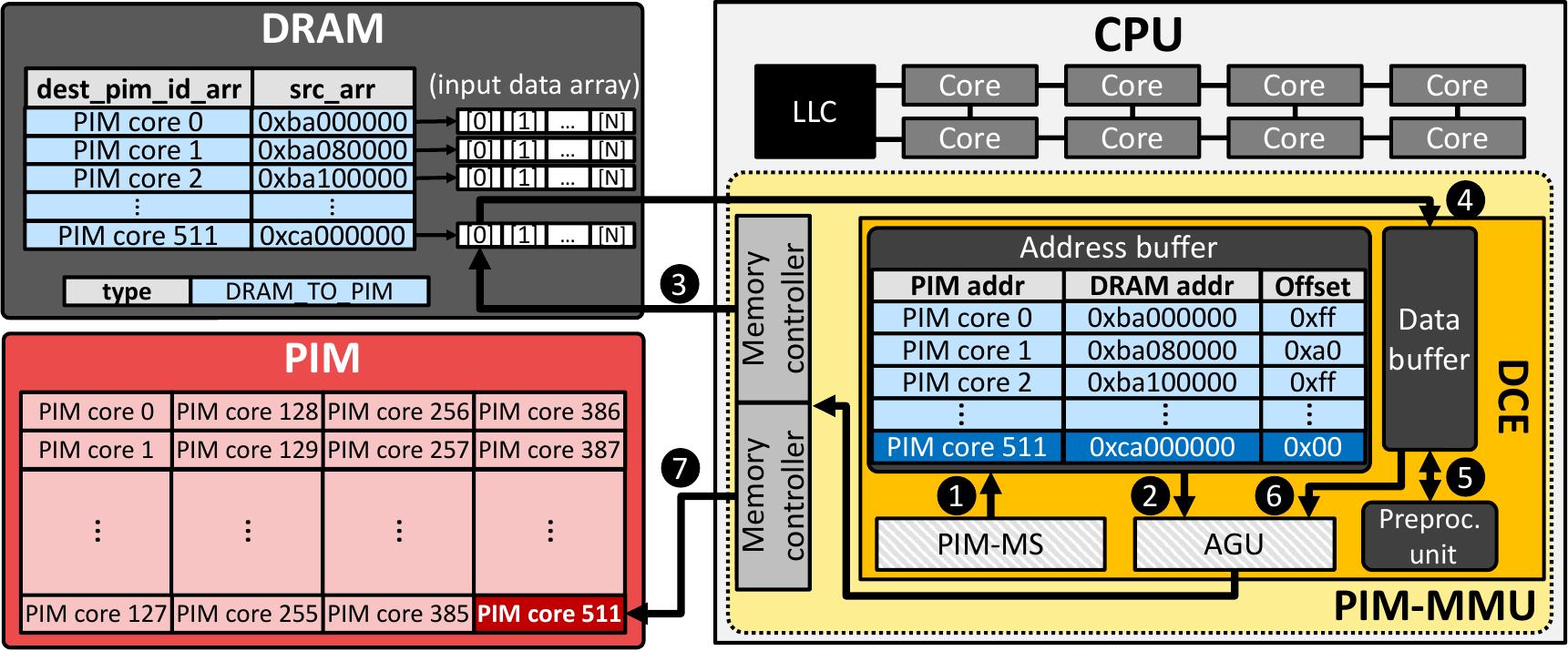} 
  \caption{Example illustrating \proposedname's overall dataflow in transferring data from DRAM to PIM.}
  \label{fig:agu_example}
	\vspace{-1.3em}
\end{figure}

\subsection{Data Copy Engine (DCE)}
\label{sect:proposed_dce}
The DCE contains the following components: (1) an Address Generation Unit
(AGU), (2) PIM-MS, (3) two SRAM-based buffers (data buffer and address buffer), and
(4) a preprocessing unit (\fig{fig:overview_proposed}).  We use the example in \fig{fig:agu_example}
to illustrate 
\proposedname's overall dataflow during a DRAM$\rightarrow$PIM data transfer
(PIM$\rightarrow$DRAM data transfer is orchestrated similarly but we omit its
 explanation for brevity). 

When the CPU launches the \texttt{pim\_mmu\_transfer} kernel for
execution, the address buffer is copied with both (1) the physical DRAM addresses (\texttt{src\_arr}) that point to all the source data
arrays (\texttt{input data array} in \fig{fig:agu_example}) and (2) the
physical PIM addresses that point to all the destination locations within PIM
(\texttt{dest\_pim\_id\_arr}) to which the source data will be written into.
Each entry in the address buffer stores the following information: (1) the base
DRAM address of the source input data array (\texttt{DRAM addr} field in
		\fig{fig:agu_example}),
				(2) the destination PIM core's ID (\texttt{PIM addr} field), and (3) an
				offset counter value (\texttt{Offset} field)
	that keeps track of the total number of data elements successfully read  from the source data.  A data transfer
operation is managed by the PIM-MS, which not only decides which memory requests to schedule to the DRAM subsystem (PIM-MS's memory scheduling algorithm and the key intuitions
		that drive its design is detailed in the following \sect{sect:proposed_pimms}), but it also goes over
the address buffer entries and coordinates the \emph{translation} of the source/destination physical addresses to DRAM/PIM addresses with the memory controller.
In the example in \fig{fig:agu_example}, we will assume
that PIM-MS has chosen to transfer data targeting PIM core ID=511. PIM-MS
first reads an entry from the address buffer (step \ding{182} in
								 \fig{fig:agu_example}) and sends it to the AGU (step \ding{183}).  The AGU then translates the source physical address
to the corresponding DRAM address with the memory controller and places the translated memory read request to the memory controller's read request queue. These two steps (\ding{182},\ding{183})
are iteratively done over all the entries in the address buffer until the memory controller's request queue is full.
When the memory controller services a read operation (step \ding{184}), the retrieved data is temporarily stored inside the data buffer (\ding{185}) and
the corresponding address buffer entry's \texttt{Offset} counter value is incremented to keep track of the data transfer progress made so far.
The returned data, stored inside the data buffer, is then read out by the preprocessing unit and is transposed on-the-fly (\ding{186}), the output of which
is sent to the AGU (\ding{187}). The AGU generates the translated, destination PIM address and places the memory write request to the memory controller's write request
queue, which eventually gets serviced by the memory controller and finalizes the DRAM$\rightarrow$PIM data transfer (\ding{188}).

\subsection{PIM-aware Memory Scheduler (PIM-MS)}
\label{sect:proposed_pimms}
\begin{figure}[t] \centering
  \includegraphics[width=0.485\textwidth]{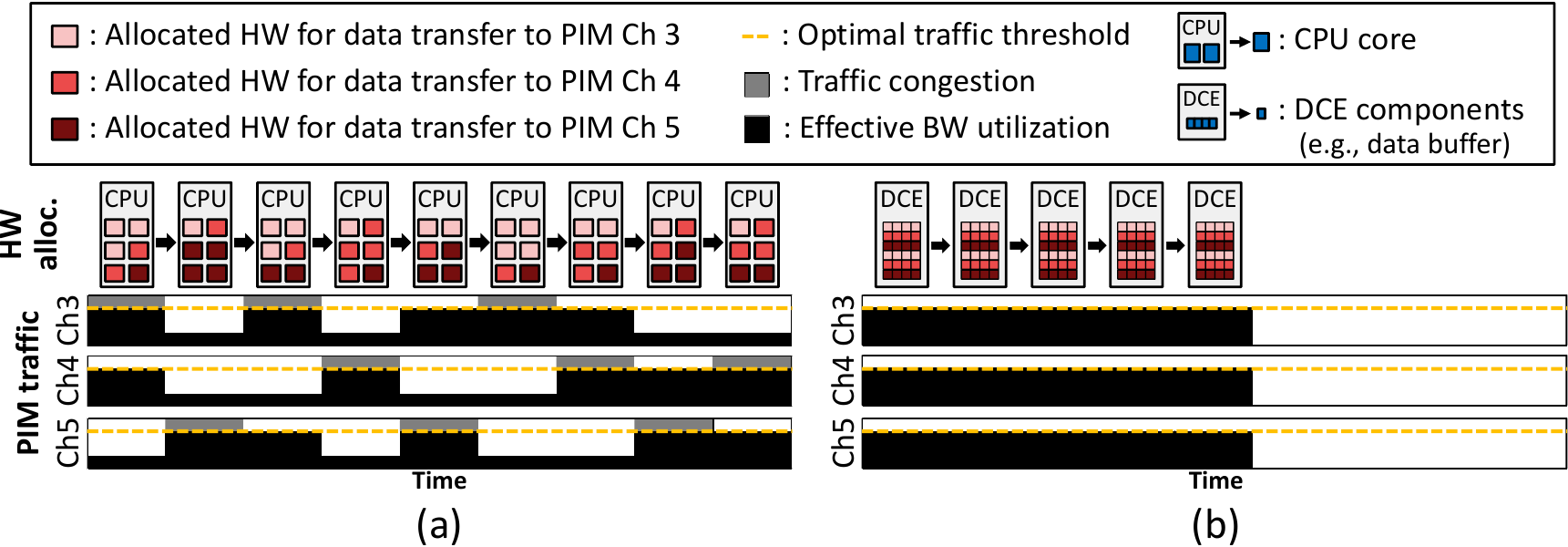} 
  \caption{Example of (a) conventional software-level/coarse-grained scheduling approach vs. (b) hardware-level/fine-grained scheduling approach of PIM-MS within DCE.}
  \label{fig:mlp_sched}
	\vspace{-1.3em}
\end{figure}

We now discuss the design principles behind our PIM-MS. The key observation
that drives PIM-MS's design is that memory transactions
targeting the PIM address space (for both reads and writes) during DRAM$\leftrightarrow$PIM
data transfer are guaranteed to have
mutually exclusive addresses. Such mutual exclusiveness ensures that there are
no true data dependencies across different PIM memory transactions. To better understand
this unique property, it is important to understand how PIM programmers
go about partitioning input/output data.  Before computation begins on the PIM
cores, the programmer partitions the input data, assigns each partition to a
specific PIM address, and transfers the partitioned data to the corresponding
PIM core.  To maintain the integrity of the offloading process and to ensure
that all input data are transferred correctly, the programmer must carefully
assign each segment of the partitioned data to a \emph{unique} PIM address.
Consequently, each segment stored within the PIM address space is mapped
independently to other segments. 

\begin{algorithm}[t]
  \caption{PIM-MS Scheduling Algorithm}
  \label{algo:pim_ms}
  \begin{algorithmic}[1]
  \scriptsize
  \rmfamily

    \STATE \textbf{Input:}  (Number of PIM cores)-sized list of tuples of (source base address, destination base address); \newline $base\_addrs$ = $[(src\_base_{0}, dst\_base_{0})\,\ldots\,(src\_base_{N}, dst\_base_{N})]$
    \STATE \textbf{Output:} List of tuples of (source address, destination address), determining the sequence in which memory transactions are scheduled;\newline
    $addrs$ = $[(src_{0}, dst_{0})\,\ldots\,(src_{N}, dst_{N})]$
    \STATE {}

    \STATE \textbf{procedure} \texttt{get\_pim\_core\_id}$(ra, bg, bk)$
    \INDSTATE \textbf{return} $ra * num\_banks * num\_bankgroups$ $+ bg * num\_banks + bk$
    \STATE \textbf{end procedure}
    \STATE {}

    \STATE \textbf{procedure} \texttt{AGU}$(id)$
    \INDSTATE {$src\_base, dst\_base = base\_addrs[id]$}
    \INDSTATE {$src\_addr = src\_base + pim\_cores[id].offset$}
    \INDSTATE {$dst\_addr = dst\_base + pim\_cores[id].offset$}
    \INDSTATE {$pim\_cores[id].offset \pluseq min\_access\_granularity$}
    \INDSTATE \textbf{return} {$src\_addr, dst\_addr$}
    \STATE \textbf{end procedure}
    \STATE {}
    
    \STATE \textbf{\#do-parallel channel}
    \STATE \textbf{begin initialization}
    \FOR {$ra \leftarrow$ $0$ to $num\_ranks$}
        \FOR{$bg \leftarrow$ $0$ to $num\_bankgroups$}
            \FOR {$bk \leftarrow$ $0$ to $num\_banks$}
                \STATE {$id = \texttt{get\_pim\_core\_id}(ra, bg, bk)$}
                \STATE {$pim\_cores[id].offset = 0$}
            \ENDFOR
        \ENDFOR
    \ENDFOR
    \STATE \textbf{end initialization}
    \STATE {}
    
    \STATE \textbf{\#do-parallel channel}
    \FOR {$bk \leftarrow$ $0$ to $num\_banks$}
        \FOR {$ra \leftarrow$ $0$ to $num\_ranks$}
            \FOR{$bg \leftarrow$ $0$ to $num\_bankgroups$}
                \STATE $id = \texttt{get\_pim\_core\_id}(ra, bg, bk)$
                \STATE {$src\_addr, dst\_addr = \texttt{AGU}(id)$}
                \STATE {$addrs.append(src\_addr, dst\_addr)$}
            \ENDFOR
        \ENDFOR
      \ENDFOR
      
  \end{algorithmic}
\end{algorithm}

With such property in mind, recall from our discussion in
\sect{sect:motivation_char} (\fig{fig:dram_vs_pim}) where we root-caused the
reason behind PIM's sub-optimal read/write throughput to the following two
factors: (1) the software-level multi-threaded DRAM$\leftrightarrow$PIM data
transfer, and (2) the fact that the OS thread scheduling policy issues data
transfer threads in coarse granularity, failing to evenly distribute read/write
traffic across the memory channels (\fig{fig:mlp_sched}(a)).  Our PIM-MS is
designed to overcome such limitation by employing a ``hardware-level''
fine-grained memory scheduling that maximizes MLP (\fig{fig:mlp_sched}(b)). As explained in
\sect{sect:proposed_sw}, upon a DRAM$\rightarrow$PIM data transfer, the \texttt{pim\_mmu\_transfer} API is invoked using a
\emph{single} thread that relays \emph{all} source and destination addresses
 to the DCE.  Therefore, when the OS schedules
	this (single) thread for execution, the 
	source and	destination physical addresses stored inside the address
	buffer gets translated into DRAM read/write requests that are targeted for \emph{all}
	destination PIM banks (unlike baseline's
			software-level/multi-threaded/coarse-grained thread scheduling where the
			memory requests available for scheduling only target a \emph{single}
			destination PIM bank at any given time). This in effect allows our PIM-MS to have
	much higher visibility and flexibility regarding
	which memory read/write requests to
	schedule to which PIM bank.

Given this opportunity, PIM-MS employs a memory scheduling algorithm
		 that exploits its enhanced visibility to maximize
		channel/bank-group/bank-level parallelism, aggressively reordering the sequence in which PIM read/write memory requests are issued to each PIM bank. We use \algo{algo:pim_ms} to describe PIM-MS's scheduling algorithm. 
  The input to \algo{algo:pim_ms} is the contents stored inside the address buffer, as detailed in \sect{sect:proposed_dce} (\fig{fig:agu_example}).  
   During initialization, the metadata representing the number of bytes to be transferred to each PIM core (i.e., \texttt{pim\_cores[id].offset}, the ``Offset'' field in the address buffer entry in \fig{fig:agu_example}) is set to 0 (line $16$-$26$).
    PIM-MS then seeks to maximize channel-level parallelism by concurrently issuing memory requests to all PIM channels (line $28$). To minimize column-to-column DRAM timing delay (\texttt{tCCD}), PIM-MS prioritizes bank group interleaving by issuing  successive column commands to access different bank groups (line $31$). 
    Lastly, using AGU translated DRAM address information, PIM-MS seeks to minimize row buffer conflicts while maximizing bank-level parallelism (line $8$-$14$).
    Overall, such hardware-level/fine-grained memory scheduling helps better utilize MLP, significantly
		 improving PIM read/write throughput vs. baseline's software-level/coarse-grained
		 memory scheduling (\fig{fig:mlp_sched}).

\subsection{Heterogeneous Memory Mapping Unit (HetMap)}
\label{sect:proposed_hetmap}

PIM manufacturers adjust the memory mapping function to prevent conflicts
between DRAM and PIM, which inevitably leads to decreased DRAM read/write
throughput (\fig{fig:pim_mapping_bw}, \sect{sect:motivation_char}).  To achieve
the dual goals of preserving high DRAM throughput while also separating the
physical address space for DRAM and PIM, we introduce a unique memory mapping
strategy called \emph{HetMap}. Illustrated on the right side of
\fig{fig:overview_proposed}, HetMap employs two separate memory mapping
functions, each optimized for a different design objective: the physical
address space reserved for PIM utilizes a \emph{locality-centric mapping}
(\fig{fig:pim_mapping_detail}(a)) whereas the physical address space allocated
for DRAM employs an \emph{MLP-centric mapping}
(\fig{fig:pim_mapping_detail}(b)).  Depending on what the physical address the
incoming memory request is targeted for, HetMap dynamically determines whether
the memory request falls within the address space of DRAM or PIM. If the memory
request is targeted for the DRAM space, it is mapped using the MLP-centric
mapping, which incorporates MLP-enhancing optimizations, i.e., XOR hashing and
placing channel bits near the LSB.  If the memory request belongs to the PIM
space, the locality-centric mapping is employed which adopts a simpler memory
mapping strategy, i.e., the order in which the DRAM hierarchy is laid out is
preserved in the locality-centric mapping (referred to as \texttt{ChRaBgBkRoCo}
		mapping in the remainder of this paper).  For example, starting from the
MSB of the physical address space, channel bits (\texttt{Ch}) are mapped first,
		followed by rank (\texttt{Ra}), bank-group (\texttt{Bg}), bank
		(\texttt{Bk}), row (\texttt{Ro}), and finally column (\texttt{Co}). As the
		sub-optimal PIM read/write throughput observed in conventional PIM devices
		is due to the software-level/coarse-grained/multi-threaded data transfers
		(a limitation which our PIM-MS effectively addresses), HetMap's
		locality-centric mapping does not degrade PIM throughput, a property we
		quantitatively demonstrate in \sect{sect:evaluation}.

In terms of implementation complexity, HetMap is co-designed with the
BIOS firmware and hardware microarchitecture as follows. During system
bootstrapping, the BIOS identifies the memory system configuration (number of
		channels/ranks/$\ldots$) as well as the total memory capacity available in
both DRAM DIMMs and PIM DIMMs. After the memory configuration is identified,
			 the BIOS firmware informs the CPU's memory controller the
			 range in which the physical address space is partitioned across DRAM vs.
			 PIM. Following this procedure, the separate address
			 mappings are established for DRAM and PIM which HetMap utilizes to
			 enforce the locality-centric and MLP-centric mapping for PIM and DRAM
			 access requests, respectively.

\subsection{\proposedname vs. Conventional DMA Engines} 

	To alleviate the
performance overhead of memory copy operations (e.g., \texttt{memcpy},
		\texttt{memmove}), there exists several Direct Memory Access (DMA)
	engines that orchestrate data copy without CPU's intervention, e.g.,  Intel
	I/OAT~\cite{ioat_sosp, ioat0, ioat1, ioat_infocom}, Intel
	DSA~\cite{intel_dsa, namsung_dsa}, and AMD PTDMA~\cite{amd_ptdma}. One might
	wonder whether  the challenges of DRAM$\leftrightarrow$PIM data transfers can
	be effectively addressed by utilizing existing DMA engines.  However, memory
	bus integrated PIM systems have fundamental architectural differences
	compared to conventional systems without PIM, limiting the efficacy of
	these DMA engines. As discussed in \sect{sect:characterization}, data
	partitioning as well as transferring partitioned data in/out of PIM is at the programmer's
	discretion, unlike conventional DRAM-only memory systems where the
	allocation of data, its partitioning, and its transfers are transparently
	handled at the hardware-level to maximize MLP.
	Consequently, current PIM systems rely on coarse-grained, multi-threaded data
	transfers to maximize MLP at the software-level. Because DMA engines are
	not designed to utilize this unique property of PIM systems for performance
	optimizations, they are not able to fully reap out the abundant parallelism
	inherent in DRAM$\leftrightarrow$PIM data transfers. As discussed in
	\sect{sect:proposed_pimms}, \proposedname's DCE and PIM-MS can
	collaboratively utilize such opportunity with our	hardware/software
	co-design, maximizing memory bandwidth utilization for both PIM read and write
	operations.

Overall, while some of the functionalities provided with DCE (especially its
		ability to independently orchestrate data transfers without the CPU's assistance)
		does resemble those of existing DMA engines, the features provided with our \proposedname is
		far beyond what current DMAs are capable of providing, e.g., the fine-grained
		scheduling of PIM-MS and HetMap's dual-mapping function.
		In \sect{sect:evaluation}, we quantitatively
		demonstrate the limitations of existing DMA engines vs. our \proposedname architecture.
\section{Methodology}
\label{sect:methodology}

\begin{table}[t!] \centering
\centering
\hfill
\caption{Baseline system and \proposedname configuration.}
\label{tab:sim_config}
\resizebox{\linewidth}{!}{
\begin{tabular}{ll}
\hline
\hline
\multicolumn{2}{c}{\textbf{Host Processor}} \\
\hline
\textbf{CPU}                                                   & \begin{tabular}[c]{@{}l@{}} 8 core, 3.2GHz, 4-wide Out-of-Order,\\ 224 entry instruction window, 64 MSHRs per core\end{tabular}  \\ \hline
\textbf{Last Level Cache (LLC)}                                                        & \begin{tabular}[c]{@{}l@{}} 8MB shared, 64B cacheline, 16-way associative\end{tabular}  \\ \hline
\textbf{Memory Controller}                                                        & \begin{tabular}[c]{@{}l@{}} 64-entry read \& write request queues, FR-FCFS, \\ locality-centric memory mapping\end{tabular}  \\ 
\hline
\hline
\multicolumn{2}{c}{\textbf{DRAM System}} \\
\hline
\textbf{Timing Parameter}                                                   & \begin{tabular}[c]{@{}l@{}} DDR4-2400\end{tabular}  \\ \hline
\textbf{System Configuration}                                                        & \begin{tabular}[c]{@{}l@{}} 4 channels, 2 ranks per channel\end{tabular}  \\
\hline
\hline
\multicolumn{2}{c}{\textbf{PIM System}} \\
\hline
\textbf{Timing Parameter}                                                   & \begin{tabular}[c]{@{}l@{}} DDR4-2400\end{tabular}  \\ \hline
\textbf{System Configuration}                                                        & \begin{tabular}[c]{@{}l@{}} 4 channels, 2 ranks per channel (512 PIM cores) \end{tabular}  \\ \hline
\hline
\multicolumn{2}{c}{\textbf{PIM-MMU}} \\
\hline
\textbf{DCE}                                                   & \begin{tabular}[c]{@{}l@{}} 3.2GHz clock frequency,\\ 16 KB data buffer, 64 KB address buffer\end{tabular}  \\ \hline
\textbf{PIM-MS}                                                        & \begin{tabular}[c]{@{}l@{}} 
Detailed in \algo{algo:pim_ms} \end{tabular}  \\ \hline
\textbf{HetMap}                                                        & \begin{tabular}[c]{@{}l@{}}(DRAM side): MLP-centric memory mapping \\ (PIM side): ChRaBgBkRoCo\end{tabular}  \\ \hline
\hline
\end{tabular}
}
\vspace{-1.3em}
\end{table}

The system characterization in \sect{sect:characterization} is conducted using a real
UPMEM-PIM system, containing an Intel Xeon Gold 5222
CPU attached with 3 channels of DDR4-3200 DIMM (total bandwidth of 76.8 GB/s)
	and 3 channels of DDR4-2400 based UPMEM-PIM DIMM (total bandwidth of 57.6
			GB/s) with one DIMM per each channel.  To demonstrate the effectiveness
	of \proposedname in \sect{sect:evaluation}, we employ a hybrid
	evaluation methodology that utilizes both cycle-level simulation and
	wall-clock time measurements from our real UPMEM-PIM system as follows.

\textbf{Simulation framework.}  Since our proposed \proposedname is designed to
improve the performance of DRAM$\leftrightarrow$PIM data transfers and not the
performance of executing the PIM kernels itself, we measure the PIM kernel execution time
using our real UPMEM-PIM server.  When estimating the
performance and energy-efficiency of DRAM$\leftrightarrow$PIM data transfers
over both baseline UPMEM-PIM and our proposed \proposedname, we utilize
cycle-level simulation by extending Ramulator~\cite{ramulator} (the
		configuration of the host CPU and its DRAM/PIM memory system is summarized in
		\tab{tab:sim_config}). To simulate the cycle count of baseline UPMEM-PIM's
data transfers without \proposedname, we compile UPMEM-PIM runtime library's
\texttt{dpu\_push\_xfer} function (one that handles DRAM$\leftrightarrow$PIM
		transfers) using gcc $9.4.0$ and extract the instruction traces that
will be executed by the CPU,
one which Ramulator's CPU-trace driven mode executes to
evaluate the DRAM$\leftrightarrow$PIM data transfer time.  
Note that the CPU model in Ramulator currently does not support AVX instructions, which are utilized by UPMEM-PIM to orchestrate DRAM$\leftrightarrow$PIM transfers. To model the effect of AVX vector load (PIM read) and store (PIM writes) instructions in our evaluation, we modified Ramulator and its core model and emulate the behavior of AVX load/store instructions by executing them as ``wide'' 64B read (PIM read) and 64B write (PIM write) memory accesses. Because memory requests targeting the PIM address space are non-cacheable, we implement these 64B read/write operations to bypass the cache, unlike the normal, cacheable (8B) read/write operations which target the DRAM address space.
To model the effect
of OS thread scheduling on baseline UPMEM-PIM's multi-threaded data transfers
(\fig{fig:dram_vs_pim}), we configure Ramulator to concurrently execute 8
data transfer operations targeting 8 PIM cores (i.e., the baseline CPU
		contains 8 cores, \tab{tab:sim_config}) which are preempted every 1.5 ms
based on a round-robin thread scheduling
policy~\cite{linux_cfs}.  Our \proposedname's
simulation time is evaluated by properly modeling the cycle-level behavior of DCE, PIM-MS, and
HetMap inside Ramulator. 

\textbf{Evaluated workloads.} The evaluation in \sect{sect:evaluation} is divided into
two parts: (1) microbenchmarks that strictly focus on 
evaluating the effect of \proposedname on DRAM$\leftrightarrow$PIM data transfer, and (2)
	real-world PIM benchmarks that evaluate \proposedname's effect on end-to-end
	performance improvement.  As for the microbenchmarks, we establish two data
	transfer workloads as follows.  First, to measure the data transfer
	throughput of DRAM$\leftrightarrow$PIM, we use the microbenchmark provided in
	the open-source PrIM~\cite{prim} benchmark suite (named \texttt{CPU-DPU}). Second, to
	measure the data transfer throughput of DRAM$\leftrightarrow$DRAM (i.e.,
			\texttt{memcpy}), we design a custom microbenchmark which employs
	multi-threading to transfer data using AVX-512 vector instructions (e.g.,
			\texttt{\_mm512\_stream\_si512}) to maximize the throughput of DRAM.
	As for the real-world PIM benchmarks, we utilize the $16$ memory-intensive
workloads from PrIM~\cite{prim}. 

\textbf{Energy and area overhead estimation.} 
The design overhead of \proposedname is primarily dominated by the $16$ KB (data buffer) and $64$ KB (address buffer) of SRAM buffers provisioned
within the DCE.
To estimate \mbox{\proposedname}'s energy consumption and area overhead on top of the
baseline and proposed system, we utilize McPAT \mbox{\cite{mcpat}} and
CACTI \mbox{\cite{cacti}} under 32nm CMOS technology, respectively.

\section{Evaluation}
\label{sect:evaluation}
All results presented
in \sect{sect:eval_microbenchmark} are based on cycle-level simulation whereas
evaluations conducted in \sect{sect:eval_benchmark} employ a hybrid of simulation
augmented with wall-clock
time measurements over a real UPMEM-PIM system (\sect{sect:methodology} details
		our methodology).

\subsection{Microbenchmarks}
\label{sect:eval_microbenchmark}
\begin{figure}[t!] \centering
%\vspace{-.3em}
  \includegraphics[width=0.47\textwidth]{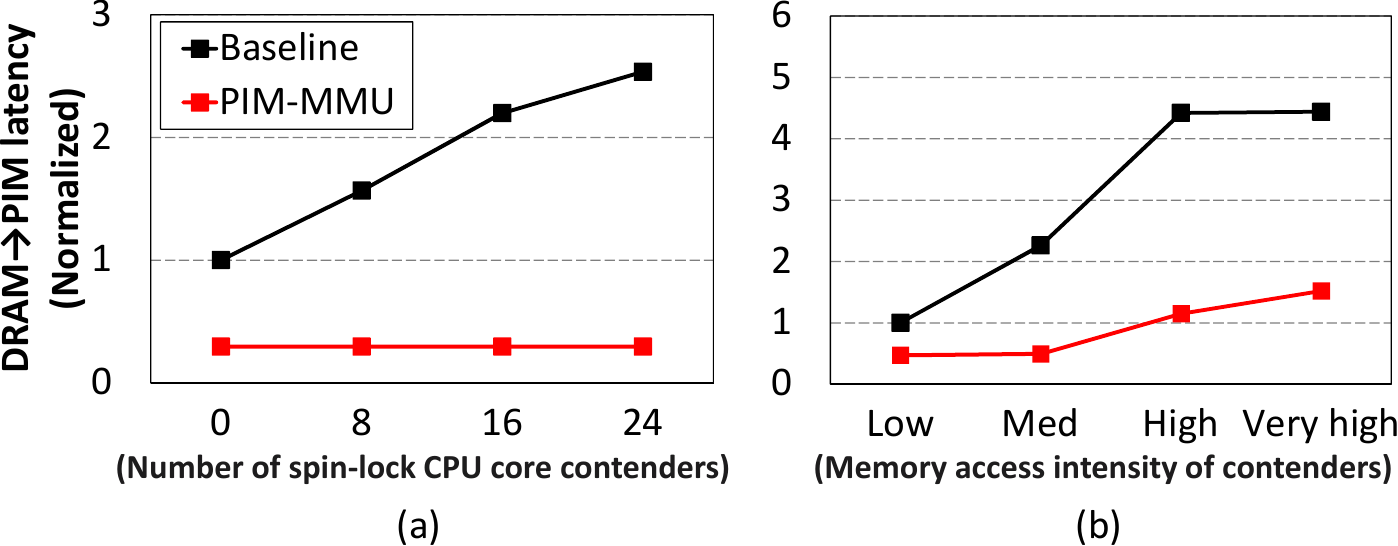}
  \caption{
  Performance sensitivity of DRAM$\rightarrow$PIM data transfer operation when it is co-located with 
  (a) compute-intensive and (b) memory-intensive contender workloads. We only present results for DRAM$\rightarrow$PIM data transfer because similar trends were observed for PIM→DRAM data transfer.
  }
  \label{fig:eval_contention}
\vspace{-1.3em}
\end{figure}

Here we evaluate \proposedname's effectiveness in improving DRAM$\leftrightarrow$PIM
data transfer performance, demonstrating:
(1) the importance of offloading DRAM$\leftrightarrow$PIM data transfers
to our DCE when \emph{other} CPU-side contenders fight over CPU compute and memory resources,
	 (2) the improvement in DRAM read/write throughput, and finally (3) the importance
	 of enhancing PIM read/write throughput via an ablation study.

\textbf{Resource contention with co-located workloads.}
In real systems, multiple workloads are typically
co-located within the same server, sharing CPU compute and memory resources. In
\fig{fig:eval_contention}, we evaluate the sensitivity of
\mbox{\proposedname} vs. baseline UPMEM-PIM's DRAM$\rightarrow$PIM data transfer performance when it is co-located with (a) compute-intensive  and (b) memory-intensive workloads.
For the co-located compute-intensive workload, we instantiate an increasing number of
spinlock-like CPU core contenders (each contender's memory accesses are primarily captured at its on-chip
caches, exhibiting compute-boundedness) that concurrently execute with
DRAM$\rightarrow$PIM data transfers (\fig{fig:eval_contention}(a)). As for the co-located memory-intensive workload, we allocate half of the CPU cores to run the resource contending workload, each contender gradually increased with higher memory access intensity (from ``low'' to ``very high'' intensity, which we tune by gradually increasing the ratio of memory instructions vs. non-memory instructions) to increasingly stress the memory subsystem and thus directly interfere with the concurrently running DRAM$\rightarrow$PIM data transfers (\fig{fig:eval_contention}(b)).

In our experiment in \fig{fig:eval_contention}(a), with an increasing number of compute-intensive CPU core contenders, the baseline system's data transfer latency sharply increases. Such performance degradation is due to the baseline's multi-threaded
data transfer implementation (which require multiple CPU
cores to achieve high performance), experiencing resource
contention with the CPU-side contenders and leading to a
reduction in the average number of CPU threads it can leverage
for data transfer operations. Our proposed PIM-MMU, on
the other hand, is virtually insensitive to the level of CPU-side resource contention as the entire data transfer process is
offloaded to our DCE, exhibiting high robustness.

When the DRAM$\rightarrow$PIM data transfer operation is co-located with memory-intensive workloads (\fig{fig:eval_contention}(b)), both PIM-MMU and  baseline suffer from aggravated performance. This is because of the memory bandwidth contention both of these design points experience, with higher (lower) resource contention occurring when the memory-intensive contender exhibits higher (lower) memory access intensity.
Nonetheless, PIM-MMU is able to achieve consistently higher performance than baseline as PIM-MMU's data transfer operation does not require any CPU compute resources. Under the baseline system, on the other hand, the data transfer process is orchestrated using CPU threads which fight over CPU cores with memory-intensive contenders, resulting in higher performance loss.

\begin{figure}[t!] \centering
%\vspace{-.3em}
  \includegraphics[width=0.47\textwidth]{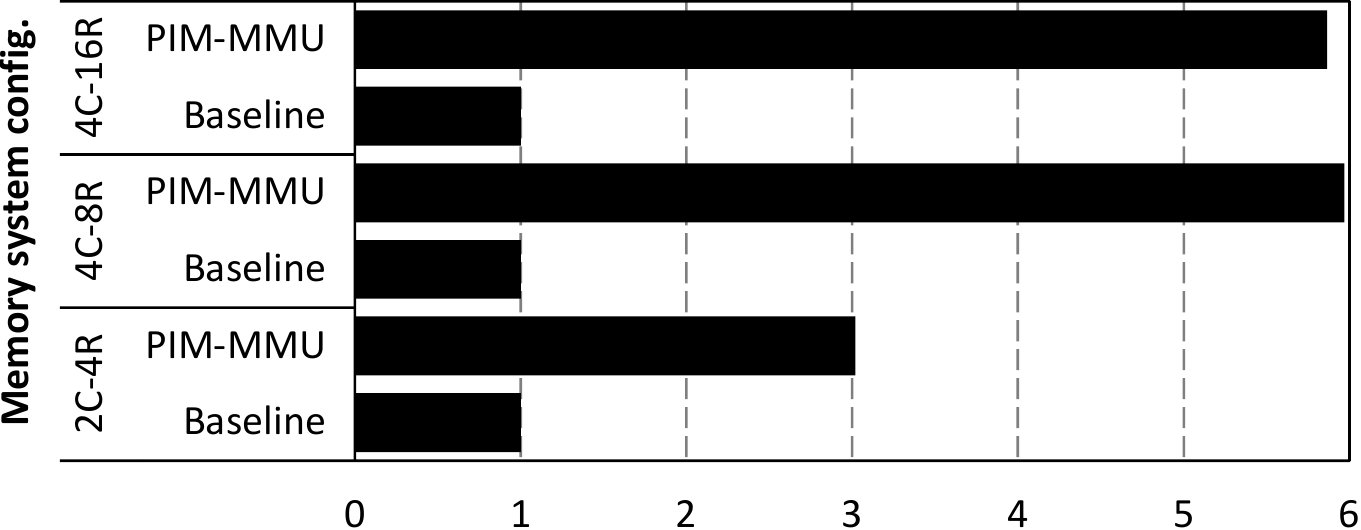}
  \caption{Normalized DRAM throughput (x-axis) during DRAM$\rightarrow$DRAM data copy operation. A `xC-yR' system configuration corresponds to a memory system containing x {\bf C}hannels and y {\bf R}anks (e.g., `2C-4R' refers to a setup with two channels and four ranks).}
  \label{fig:eval_dram_bw}
  \vspace{-1.3em}
\end{figure}
\begin{figure*}[t!] \centering
%\vspace{-1.3em}
\includegraphics[width=0.95\textwidth]{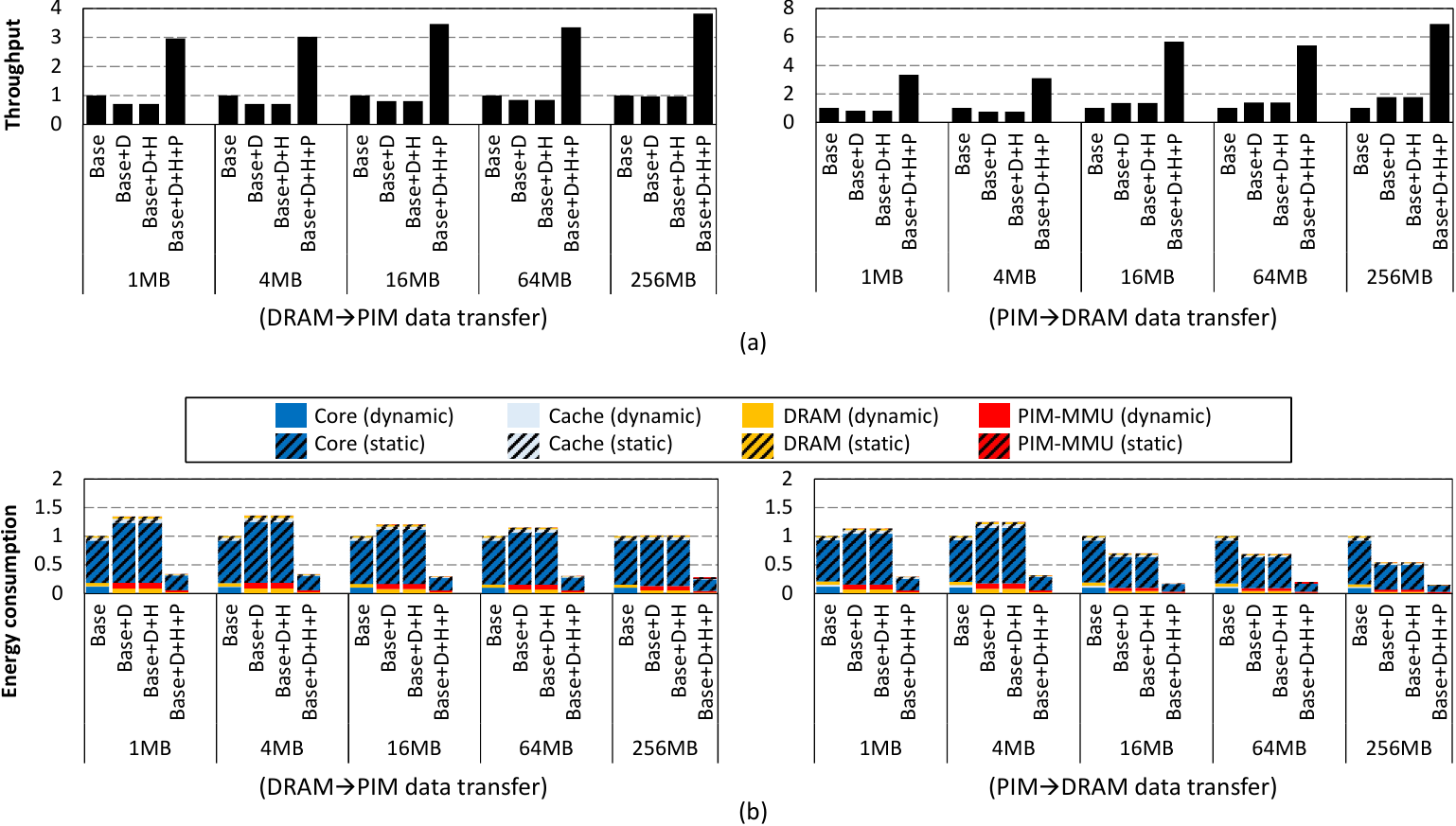}
    \caption{Ablation study to explore the effectiveness of \mbox{\proposedname}'s three key features in (a) improving the data transfer throughput and (b) reducing the chip-wide energy consumption
    during DRAM$\rightarrow$PIM and PIM$\rightarrow$DRAM data transfers with different data transfer sizes (x-axis).     
    We incrementally add (D) DCE that does not utilize PIM-MS, (H) HetMap, and (P) PIM-MS on top of the baseline system (Base) to evaluate \mbox{\proposedname}'s effectiveness. }
  \vspace{-0.7em}
	\label{fig:bw_energy_eval}
\end{figure*}
\textbf{DRAM throughput.} In \fig{fig:eval_dram_bw}, we show the DRAM
throughput during DRAM-to-DRAM data transfers (\texttt{memcpy}) as means to
demonstrate how well \proposedname's HetMap unlocks the MLP available in normal
DRAM channels.  Overall, \proposedname consistently outperforms baseline,
					 achieving an average throughput improvement of $4.9\times$ (maximum
							 $6.0\times$).  This significant increase in DRAM throughput is
					 enabled by our HetMap which not only supports separate address
					 spaces for DRAM and PIM but also facilitates MLP-centric mapping
					 just for the DRAM address space. Because MLP-centric mapping
					 effectively leverages channel-level parallelism, \proposedname's
					 DRAM throughput increases linearly with the number of channels. It
					 is important to note that, when the number of ranks increases, DRAM
					 throughput does not increase correspondingly because adding more
					 ranks only helps increase memory capacity but not memory bandwidth.

\textbf{Ablation study.} We summarize our ablation study that quantifies how much the baseline system's (denoted ``Base'')  DRAM$\leftrightarrow$PIM
data transfer throughput (\fig{fig:bw_energy_eval}(a)) and energy-efficiency (\fig{fig:bw_energy_eval}(b)) can improve by adding \proposedname's key features in an additive manner: (\textbf{D}) DCE that does not utilize PIM-MS, (\textbf{H}) HetMap, and (\textbf{P}) PIM-MS. 

Starting with
the ``Base+D'' design (i.e., baseline system utilizing DCE's DMA capability but \emph{without}
the hardware-level/fine-grained memory scheduling enabled with PIM-MS),
		this design point functions
as a proxy for  conventional DMA engines like Intel's I/OAT~\cite{ioat_sosp} or DSA~\cite{intel_dsa}.
Interestingly, ``Base+D'' actually incurs a degradation in data transfer throughput
for $7$ out of the $10$ experiments we conduct in \fig{fig:bw_energy_eval}(a).
Careful analysis of such phenomenon reveals that, compared to ``Base+D'' (i.e., the vanilla DCE that does
not employ HetMap and PIM-MS), the baseline system
that does not utilize DMA actually does a better job in utilizing memory bandwidth
thanks to its AVX-512 based wide vector read/write requests which are aggressively issued 
concurrently using the out-of-order execution cores.
With the addition of HetMap, the ``Base+D+H'' design point is able to significantly
improve the DRAM read/write throughput (as demonstrated through our microbenchmark study in \fig{fig:eval_dram_bw}), but the improvement in end-to-end DRAM$\leftrightarrow$PIM performance is still 
marginal. This is because the performance of ``Base+D+H'' gets bottlenecked on the low PIM read/write throughput as it is still based on a software-level/coarse-grained data transfer, throttling
the level of MLP it can exploit.
Once PIM-MS is employed, however, the ``Base+D+H+P'' design (i.e., PIM-MMU) fully unlocks the PIM read/write throughput
and significantly improves the performance of DRAM$\leftrightarrow$PIM transfers.

When it comes to energy-efficiency, the energy consumed by the processor-side components dominates the system-wide energy consumption (\mbox{\fig{fig:bw_energy_eval}(b)}). Consequently, the overall energy-efficiency is determined by how long it takes to finalize the DRAM$\leftrightarrow$PIM data transfer operations. Because both ``Base+D'' and ``Base+D+H'' experience longer data transfer time, these two data points suffer from higher energy consumption than ``Base''. In contrast, with all three of our  proposals in place (``Base+D+H+P''), PIM-MMU can significantly reduce data transfer latency which directly translates into lower energy consumption, achieving an average $3.3\times$ (max $3.8\times$) and $4.9\times$ (max $6.9\times$) higher energy-efficiency for DRAM$\rightarrow$PIM and PIM$\rightarrow$DRAM transfers, respectively. 

\begin{figure*}[t!] \centering
%\vspace{-.3em}
\includegraphics[width=0.95\textwidth]{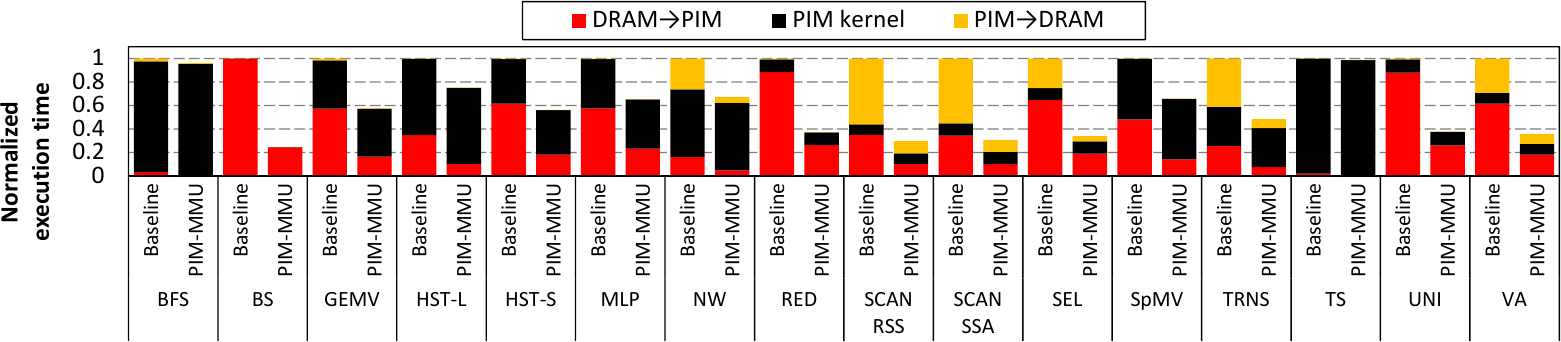}
    \caption{Normalized end-to-end execution time of PIM benchmarks. PIM kernel execution time is measured using our real UPMEM-PIM system while DRAM$\leftrightarrow$PIM data transfer time is properly scaled between baseline vs. \proposedname based on our simulation results.} 
  \vspace{-1.3em}
	\label{fig:evaluation_prim}
\end{figure*}
\subsection{Real-World PIM Benchmarks}
\label{sect:eval_benchmark}

\fig{fig:evaluation_prim} shows the normalized execution time of $16$ memory-intensive
PIM workloads from PrIM~\cite{prim}. As shown, 
the latency to transfer input/output data across the DRAM and PIM address space
incurs non-trivial performance overhead, accounting to as much as $99.7\%$ (average $63.7\%$) of end-to-end execution time and underscoring the importance of resolving this
system-level bottleneck. 
\proposedname provides an 
average $3.3\times$ (max $4.1\times$) and  
an average $3.8\times$ (max $5.7\times$) reduction in DRAM$\rightarrow$PIM and PIM$\rightarrow$DRAM
data transfer latency, respectively.
The level of end-to-end performance improvement \proposedname provides is obviously dependent
on how critical the DRAM$\leftrightarrow$PIM data transfer is, e.g., TS shows marginal performance improvement with our proposed system since data transfer is not a bottleneck.
Nonetheless, 
\proposedname provides an average $2.2\times$ (max
		$4.0\times$) improvement in end-to-end performance across the entire PrIM
benchmark suite, justifying its adoption in memory bus integrated PIM systems.

\subsection{Implementation Overhead}
\label{sect:eval_overhead}

Implementation of PIM-MS and HetMap is primarily dominated by logic gates, so
\proposedname's most significant area overhead comes from
the DCE's SRAM buffers whose size is $16$ KB and $64$ KB for
data buffer and address buffer, respectively. The area overhead of these buffers
are evaluated as $0.85$ mm$^2$ using CACTI~\cite{cacti} which amounts to only
a $0.37\%$ increase in CPU die size. Given the significant 
energy-efficiency improvement \proposedname provides, we believe such implementation 
overhead is reasonable.
\section{Related Work}
\label{sect:related}

\textbf{Data movement support.}
To alleviate CPU's burden during data movement operations (e.g.,
		\texttt{memcpy}), there exists 
a variety of DMA engines. These include
	Intel's I/OAT~\cite{ioat_sosp, ioat0, ioat1, ioat_infocom} and DSA~\cite{intel_dsa, namsung_dsa}, along with AMD's PTDMA~\cite{amd_ptdma}. Kuper et al.~\cite{namsung_dsa}, for instance, highlights the efficiency of
	 Intel's DSA in offloading DRAM$\leftrightarrow$DRAM data transfers. Additionally, NVIDIA's H100 GPU~\cite{hc_h100} introduced
	 a memory copy engine called Tensor
			 Memory Accelerator which efficiently transfers data between off-chip DRAM and on-chip scratchpad. In general, the concept
	 of offloading a data copy operation to a dedicated data movement accelerator is similar
	 between \proposedname and these DMA engines.  However, as we
	 quantitatively demonstrated through our ablation study in \sect{sect:eval_microbenchmark} (\fig{fig:bw_energy_eval},
 existing DMA engines are not optimized to exploit the unique
	 characteristics of PIM let alone its implication from a system's perspective, failing to
	 fully reap out MLP to accelerate DRAM$\leftrightarrow$PIM data transfers. There also exists several prior work proposing architectures for bulk data transfer acceleration~\cite{seshadri2015fast, seshadri2016buddy, rowclone, simdram, ambit, lisa, pluto}. RowClone~\cite{rowclone} and SIMDRAM~\cite{simdram}, for instance, proposes an in-DRAM bulk data transfer acceleration scheme where multiple
 DRAM rows are concurrently activated as means to enable fast row to row data copies. These solutions, however, can only copy data between rows that reside within the same DRAM chip, rendering \proposedname's contribution orthogonal these prior work as we focus on accelerating DRAM$\leftrightarrow$PIM data transfers. These results highlight
	 the unique
	 contribution and novelty of \proposedname vs. conventional DMA engines or bulk data transfer architectures. 

\textbf{Characterization of commercial PIM device.} Several recent work
conducted a detailed characterization of commercial PIM systems.
For instance, \cite{prim, prim_2, prim_3} provides a detailed workload characterization
on the UPMEM-PIM system with another line of research investigating
how to exploit the UPMEM-PIM system to accelerate data-intensive workloads, e.g., 
 dense/sparse linear algebra, databases, data analytics,
				 graph processing, bioinformatics, image processing, compression,
				 simulation, and encryption
				 \cite{dna_mapping_using_processor_in_memory_architecture,
					 upmem_sigmod, bulk_jpeg_decoding_on_in_memory_processors,
					 a_case_study_of_processing_in_memory_in_off_the_shelf_systems,
					 homomorphic_upmem, high-throughput_pairwise_upmem, trans_pim_lib,
					 pim_tree, uppipe, dnn_in_upmem, adaptive_query_compilation_upmem,
					 pimDB, jk_sigmetric}.
There also exists a series of studies exploring the hardware/software architectural 
support for Samsung's HBM-PIM architecture~\cite{hbm_pim_isscc, hbm_pim_isca}, as well as
studies on using Samsung's near-memory processing based AxDIMM for accelerating
recommendation models \cite{axdimm} and database operations
					 \cite{axdimm_db}. While these prior work provide invaluable insights on
					 commercial PIM devices, to the best of our knowledge, \proposedname, which is an extension of our prior work~\cite{pimmmu_cal}, is the first
					 to uncover the unique system-level challenges of DRAM$\leftrightarrow$PIM data transfers in memory bus integrated PIM systems.

\textbf{Memory management for PIM systems.} Prior work explored the designs of memory management targeting PIM systems~\cite{hall2000memory, azarkhish2016design, zhang2020meg, lazypim, near_data_acceleration_with_concurrent_host_access, conda, syncron, tom, impica, to_pim_or_not}. Hall et al.~\cite{hall2000memory} described memory management requirements associated with virtual memory, specifically for the Data IntenSive Architecture (DIVA)~\cite{diva}. Azarkhish et al.~\cite{azarkhish2016design} presented a zero-copy pointer
passing mechanism to allow low overhead data sharing between the host and PIM with virtual memory support. Zhang et al.~\cite{zhang2020meg} presented an IOMMU design that efficiently handles massive memory requests while supporting virtual memory for PIM.
While these works present memory management schemes for PIM-enabled systems, they mainly focus on aspects related to virtual memory, distinguishing \proposedname's contribution from them.

\textbf{Memory mapping.} There also exists several prior work exploring efficient memory mapping architectures ~\cite{ghasempour2016dream, gpu_mapping, xor_mapping, zhang2022software, meswani2015heterogeneous, li2017utility, dong2010simple, kaseridis2010bandwidth, chatterjee2014managing, akin2015data, kaseridis2011minimalist}. Ghasempour et al. ~\cite{ghasempour2016dream} employed multiple DRAM address mapping functions, dynamically choosing the optimal address mapping function at runtime based on the target workload's unique memory access pattern for improved DRAM performance. Zhang et al.~\cite{zhang2022software} proposed user-program behavior-aware memory mapping, which can efficiently exploit channel-level parallelism. Meswani et al.~\cite{meswani2015heterogeneous} proposed hardware/software co-designed memory management for die-stacked DRAM. Li et al.~\cite{li2017utility} introduced a hybrid memory management approach that quantitatively assesses the performance gains of migrating a page between various memory types within a hybrid memory system. While these prior work bears some similarity with \proposedname's HetMap architecture, the unique aspect of \proposedname lies in demonstrating the importance of synergistically combining DCE, PIM-MS, and HetMap to fully unlock performance, rendering our contribution unique.
\section{Conclusion}
\label{sect:conclusion}

Current PIM-integrated systems suffer from high performance overheads during 
DRAM$\leftrightarrow$PIM data transfers. We propose \proposedname, a hardware/software co-design that enables
energy-efficient data transfers in memory bus integrated PIM.
Compared to baseline, \proposedname incurs negligible implementation overheads
while providing energy-efficiency improvements in DRAM$\leftrightarrow$PIM data transfers, leading to an end-to-end $2.2\times$
speedup for real-world PIM workloads.

\section*{Acknowledgment}
This work was partly supported by the National Research Foundation of Korea (NRF) grant funded by the Korea government (MSIT) (NRF-2021R1A2C2091753), Institute of Information \& Communications Technology Planning \& Evaluation (IITP) grant funded by the Korea government (MSIT) (No.RS-2024-00438851, (SW Starlab) High-performance Privacy-preserving Machine Learning System and System Software), (No. 2022-0-01037, Development of High Performance Processing-in-Memory Technology based on DRAM), (No.RS-2024-00395134, DPU-Centric Datacenter Architecture for Next-Generation AI Devices), (No.RS-2024-00402898, Simulation-based High-speed/High-Accuracy Data Center Workload/System Analysis Platform), and IITP under the Graduate School of Artificial Intelligence Semiconductor (IITP-2024-RS-2023-00256472) grant funded by MSIT. The EDA tool was supported by the IC Design Education Center (IDEC), Korea. We also appreciate the support from Samsung Electronics. Minsoo Rhu is the corresponding author.

\bibliographystyle{IEEEtranS}
\bibliography{refs}

\end{document}